\documentclass[preprint,amsmath,nofootinbib,aps,prd,superscriptaddress,tightenlines,12pt]{revtex4}
\pdfoutput=1
\usepackage{graphicx}
\usepackage{graphics}
\usepackage{amsmath}
\usepackage{amssymb}
\usepackage{hyperref}
\usepackage{xcolor}
\usepackage{xspace}
\usepackage{subfigure}

\newcommand{\eq}[1]{Eq.~\eqref{eq:#1}}
\newcommand{\eqs}[2]{Eqs.~\eqref{eq:#1} and \eqref{eq:#2}}

\renewcommand{\sec}[1]{Sec.~\ref{sec:#1}}

\newcommand{\subsec}[1]{Sec.~\ref{subsec:#1}}

\newcommand{\fig}[1]{Fig.~\ref{fig:#1}}
\newcommand{\figs}[2]{Figs.~\ref{fig:#1} and \ref{fig:#2}}
\newcommand{\app}[1]{Appendix~\ref{app:#1}}

\newcommand{\tab}[1]{Table~\ref{tab:#1}}

\newcommand{\beq}{\begin{eqnarray}}
\newcommand{\eeq}{\end{eqnarray}}

\newcommand{\ord}[1]{\mathcal{O}\!\left(#1\right)}
\newcommand{\as}{\alpha_s}
\newcommand{\df}{\textrm{d}}
\newcommand{\Mll}{M_{\ell \ell}}
\newcommand{\Mlnu}{M_{\ell \nu}}
\newcommand{\MT}{M_T}
\newcommand{\MZ}{M_Z}
\newcommand{\MW}{M_W}
\newcommand{\Zproc}{pp \to Z^*/\gamma^* \to \ell^+\ell^-}
\newcommand{\Wproc}{pp \to W^* \to \ell\nu}
\newcommand{\GeV}{\text{GeV}}
\newcommand{\TeV}{\text{TeV}}
\newcommand{\MET}{\mbox{$E_T\hspace{-0.25in}\not\hspace{0.20in}$}}

\begin{document}

\setlength\baselineskip{17pt}

\begin{flushright}
\vbox{
\begin{tabular}{l}
\end{tabular}
}
\end{flushright}
\vspace{0.1cm}


\title{\bf Running Electroweak Couplings as a Probe of New Physics}

\vspace*{1cm}

\author{Daniele~S.~M.~Alves}
\affiliation{Center for Cosmology and Particle Physics, Department of Physics, New York University, New York, NY 10003}
\affiliation{Department of Physics, Princeton University, Princeton, NJ 08544}
\author{Jamison~Galloway}
\affiliation{Center for Cosmology and Particle Physics, Department of Physics, New York University, New York, NY 10003}
\author{Joshua~T.~Ruderman}
\affiliation{Center for Cosmology and Particle Physics, Department of Physics, New York University, New York, NY 10003}
\author{Jonathan~R.~Walsh}
\affiliation{Ernest Orlando Lawrence Berkeley National Laboratory, University of California, Berkeley, CA 94720, U.S.A.}
\affiliation{Center for Theoretical Physics, University of California, Berkeley, CA 94720, U.S.A.}

\date{\today\\ \vspace*{1cm} }

  \vspace*{0.3cm}

\begin{abstract}
  \vspace{0.5cm}

The energy dependence of the electroweak gauge couplings has not been measured above the weak scale.  We propose that percent-level measurements of the energy dependence of $\alpha_{1,2}$ can be performed now at the LHC and at future higher energy hadron colliders.  These measurements can be used to set limits on new particles with electroweak quantum numbers without relying on any assumptions about their decay properties.  The shape of the high invariant mass spectrum of Drell-Yan, $p p \rightarrow Z^*/\gamma^* \rightarrow \ell^+ \ell^-$, constrains $\alpha_{1,2}(Q)$, and the shape of the high transverse mass distribution of $p p \rightarrow W^* \rightarrow \ell \nu$ constrains $\alpha_{2}(Q)$.  We use existing data to perform the first fits to $\alpha_{1,2}$ above the weak scale.  Percent-level measurements are possible because of high precision in theoretical predictions and existing experimental measurements.  We show that the LHC already has the reach to improve upon electroweak precision tests for new particles that dominantly couple through their electroweak charges.  The 14~TeV LHC is sensitive to the predicted Standard Model (SM) running of $\alpha_2$, and can show that $\alpha_2$ decreases with energy at $2-3 \sigma$ significance. A future 100~TeV proton-proton collider will have significant reach to measure running weak couplings, with sensitivity to the SM running of $\alpha_2$ at $4-5 \sigma$ and sensitivity to winos with masses up to $\sim$ 1.3~TeV at $2\sigma$.

\end{abstract}

\maketitle

\tableofcontents

\newpage

\section{Introduction}
\label{sec:intro}

Renormalization group running of the gauge couplings is a fundamental concept in quantum field theory and an important prediction of the Standard Model (SM).  Running leads to the SM prediction of asymptotic freedom of $SU(3)_C$ and $SU(2)_W$~\cite{Gross:1973id,Politzer:1973fx} and allows for the possibility of gauge coupling unification at high energies~\cite{Georgi:1974sy,Dimopoulos:1981zb,Dimopoulos:1981yj}.  It is challenging to measure gauge coupling running, because the couplings run logarithmically, requiring precision measurements conducted across a wide range of energies.

The energy dependence of $\alpha_3(Q)$ has been constrained across three decades of energy, from $Q=m_\tau$, using $\tau$ decays, up to high energies of $Q \sim 1$~TeV, using LHC studies of jet physics (see Refs.~\cite{Chatrchyan:2013txa,CMS:2013zda,CMS:2013yua} for LHC measurements of $\alpha_3$).  Intermediate energies are also constrained using a variety of techniques including lattice QCD, deep inelastic scattering, studies of heavy quarkonia, $e^+ e^-$ production of jets, and $Z$-pole measurements~\cite{Agashe:2014kda,Bethke:2011tr}.  These measurements collectively provide spectacular confirmation of the running of $\alpha_3$ predicted in the SM, confirming asymptotic freedom over these energies.  

Measurements of running $\alpha_3$ are not simply useful as tests of the SM\@.  They also constrain the possible presence of new colored states, which if present shift the $\beta$-function of $\alpha_3$, changing its energy dependence.  For example, Ref.~\cite{Kaplan:2008pt} used event shape data at LEP, which are sensitive to the running of $\alpha_3$, to set a model-independent limit on the gluino: $M_3 > 51$~GeV at 95\% confidence level.  This remains the only model-independent limit on the gluino, because all other collider searches, such as those conducted at the LHC, rely on specific assumptions about how the gluino decays.  A recent work~\cite{Becciolini:2014lya} finds that LHC data on running $\alpha_3$ can potentially improve this limit significantly, once measurement uncertainties are properly accounted for.

Measurements of the energy dependence of the ElectroWeak (EW) couplings, $\alpha_{1,2}$, can provide another nontrivial test of the SM\@.  So far, the EW couplings have been measured to exquisite precision at the $Z$-pole: $\sin^2 \theta_W(\MZ) = 0.23126(5)$~\cite{Agashe:2014kda} and $\alpha_{EM}(\MZ)^{-1} = 128.951(45)$~\cite{Burkhardt:2011ur}.  There are also measurements of $\sin^2 \theta_W$ and $\alpha_{EM}$ at energies below $\MZ$~\cite{Agashe:2014kda,Abbiendi:2005rx,Achard:2005it}, confirming that these parameters run as predicted in the SM at low energies.  However, there presently exists no measurements of the EW couplings, $\alpha_{1,2}$, at energies above $\MZ$, where electroweak symmetry is restored.  We note that LEP2 ran up to a center of mass energy of 209 GeV, allowing for such measurements just above $\MZ$.  Hadron colliders, however, can leverage their large phase space to measure EW couplings well above the weak scale.

\begin{figure}[htb]
\begin{center}
\includegraphics[width=9.0cm]{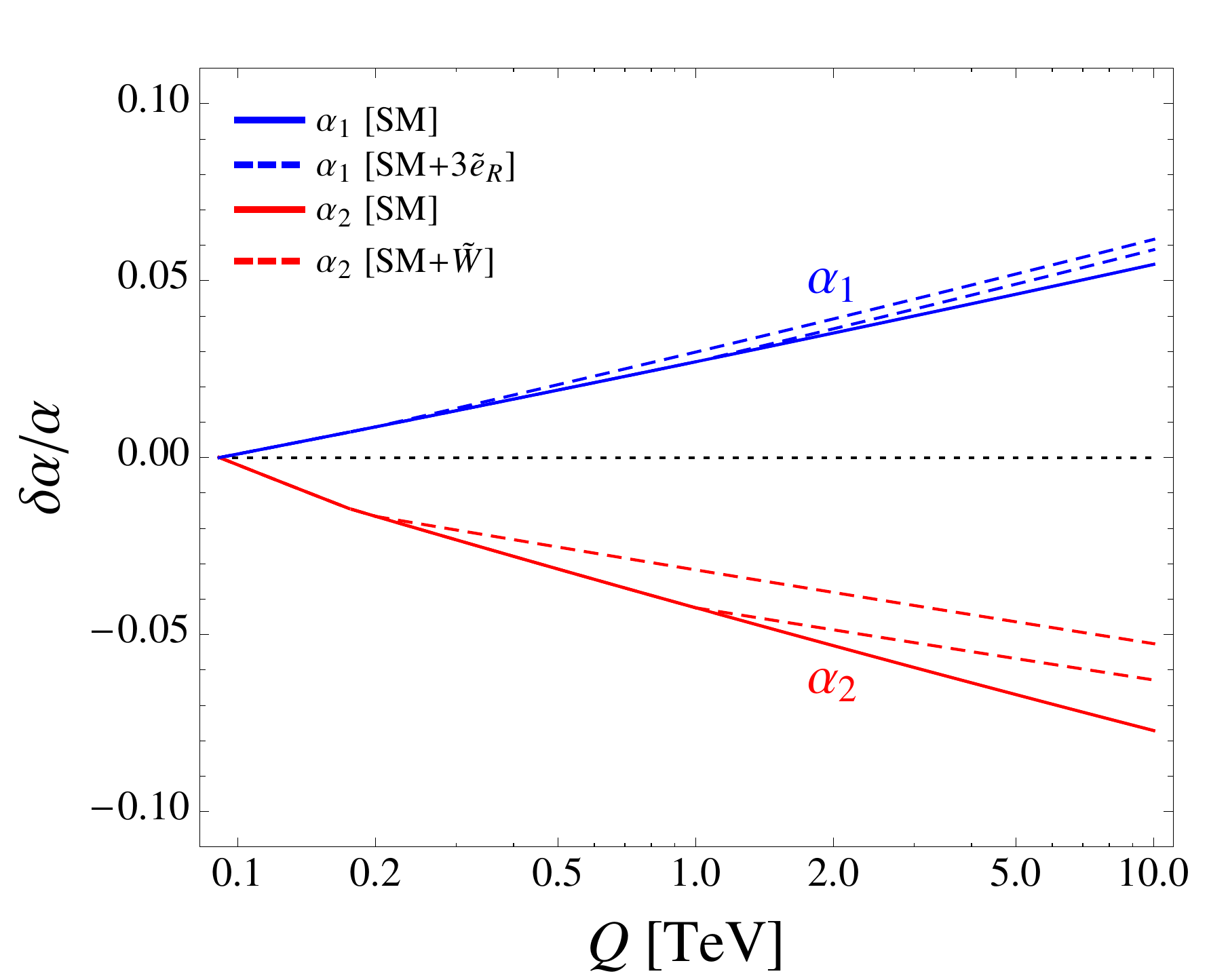}
\caption{\small {Differential running of EW couplings in the SM (solid) and in the presence of new states (dashed).  We show the fractional change in each coupling relative to its value at $\MZ$,  $\delta \alpha / \alpha = \alpha_i(Q) / \alpha_i(\MZ) - 1$.  A wino is added in the case of $SU(2)_L$, and three right-handed sleptons in the case of $U(1)_Y$; both are shown assuming masses of 200 GeV or 1 TeV\@.}}
\label{fig:run1}
\end{center}
\end{figure}

We show the SM predictions for the running of EW couplings in \fig{run1}, in terms of the fractional change of $\alpha_{1,2}$ normalized to their values at $\MZ$.    Note that in the SM, $\alpha_2$ is asymptotically free, while $\alpha_1$ gets larger with energy.  At 1 (10)~TeV, $\alpha_2$ is predicted to decrease by 3.9 (7.4)\%, and $\alpha_1$ is predicted to increase by 2.7 (5.5)\%.  If there exist new states beyond the SM with EW quantum numbers, then the energy dependence of $\alpha_{1,2}$ is modified at energies above the masses of the new states.  In \fig{run1}, we also show how the running is deflected by the presence of a wino or right-handed sleptons, with masses of 200~GeV or 1~TeV\@.   If enough new states are present, the sign of $\beta_2$ can flip and asymptotic freedom of $\alpha_2$ can be lost, as predicted in the Minimal Supersymmetric Standard Model (MSSM).

It is clear from \fig{run1} that percent-level measurements of $\alpha_{1,2}$ above $\MZ$ are desirable.  The purpose of this paper is to argue that hadron colliders, such as the LHC, can be used to measure running EW couplings.  In order to measure running EW couplings, we can compare data to theory predictions for processes with cross sections that depend on $\alpha_{1,2}$.  To achieve a high precision measurement, we would like to identify processes that meet three criteria: 
\begin{itemize}
\item[1.]  The process should have a large cross section to produce events with high momentum-transfer, where the amplitude probes the values of EW couplings at high energy.  A large cross section is necessary for small statistical uncertainties.
\item[2.]  The process should be under good theoretical control, with minimal uncertainties.  At hadron colliders, these uncertainties are usually dominated by QCD scale and PDF variation.
\item[3.] The measurement should be under good experimental control, with minimal experimental uncertainties.
\end{itemize}

In this paper, as in \cite{Rainwater:2007qa}, we focus on simple processes that meet the above criteria: the large invariant mass regime of the Drell-Yan (DY) neutral and charged current processes, $\Zproc$ and $\Wproc$.  The invariant mass of the $Z^*/\gamma^*$ sets the scale where $\alpha_{1,2}$ should be evaluated, and therefore the shape of the dilepton mass spectrum is sensitive to $\alpha_{1,2}(Q)$.  High mass $W^*$ production is sensitive to $\alpha_2(Q)$, and can be isolated using the shape of the transverse mass spectrum, since $M_{W^*} \ge \MT  $.  The Drell-Yan cross sections are under excellent theoretical control: the cross section has been computed to NNLO in QCD~\cite{Hamberg:1990np,Anastasiou:2003yy,Anastasiou:2003ds,Melnikov:2006kv,Catani:2009sm,Gavin:2010az,Li:2012wna} and NLO in EW for $Z/\gamma^*$~\cite{Berends:1984xv,Baur:1997wa,Baur:2001ze,CarloniCalame:2007cd,Arbuzov:2007db,Dittmaier:2009cr} and $W^*$~\cite{Wackeroth:1996hz,Baur:1998kt,Dittmaier:2001ay,Baur:2004ig,Arbuzov:2005dd,CarloniCalame:2006zq}.  As we will see, the scale and PDF uncertainties are of order $1\%$ in the phase space region of interest.  Drell-Yan is also amenable to high-precision experimental measurements because it depends on final state leptons whose momenta can be measured accurately.  High invariant mass $Z^*/\gamma^*$ has been measured by ATLAS at 7 TeV~\cite{Aad:2013iua}, and by CMS at 7 TeV~\cite{Chatrchyan:2013tia} and 8 TeV~\cite{CMS:2014hga}.  These measurements have already achieved high precision, with $\lesssim1-2\%$ uncorrelated uncertainties among different invariant mass bins\footnote{Fully correlated uncertainties, such as from the luminosity and lepton identification efficiency, are not important since the {\it shape} of the spectrum can be used to constrain running $\alpha_{1,2}(Q)$.}.

Precision measurements of the running of EW couplings will allow for the first model-independent 
constraints on new particles with EW quantum numbers at a hadron collider, analogous to the model-independent constraint on the gluino~\cite{Kaplan:2008pt} discussed above.  New states with EW quantum numbers are highly motivated because they are predicted in models that naturally address the hierarchy problem and in models of dark matter~\cite{Cirelli:2005uq,Cirelli:2007xd}.  Existing collider constraints are often weak due to the small production cross section, relative to colored states, and always depend on assumptions about how the new states decay.  Constraints from running couplings are complementary to other indirect constraints, such as electroweak precision tests (see {\it e.g.} Ref.~\cite{Barbieri:2004qk}) and measurements of Higgs couplings (see {\it e.g.} Ref.~\cite{Carmi:2012yp,Azatov:2012bz,Espinosa:2012ir}).

In order to discover or constrain new physics, it is desirable to measure running couplings at as high energies as possible.  Therefore, the increase of LHC energy to $13-14$~TeV will provide an important jump in reach.  One goal of this paper is to assess the reach of the 14~TeV LHC to measure running couplings.  Recently, planning has begun for a possible future proton-proton collider at higher energy, such as $\sqrt s = 100$~TeV\@.  There are preliminary proposals to host such a collider at CERN~\cite{CernFutureCollider} or in China~\cite{ChineseCollider}.  In this paper, we explore the reach of a future 100 TeV collider to measure running EW couplings.  For other recent studies of 100 TeV collider phenomenology, see {\it e.g.} Refs.~\cite{Zhou:2013raa,Cohen:2013xda,Jung:2013zya,Rizzo:2014xma,Low:2014cba,Cohen:2014hxa,Larkoski:2014bia,Hook:2014rka,Cirelli:2014dsa,Curtin:2014jma,Acharya:2014pua,Gori:2014oua}.

Several previous works have studied the impact of supersymmetry on Drell-Yan production.  Ref.~\cite{Rainwater:2007qa} suggests using Drell-Yan production at the LHC to differentiate SM versus supersymmetric running of EW gauge couplings.  Their analysis studies the reach of making a single cut on invariant mass or transverse mass, and uses simple estimates of systematic uncertainties.  We improve upon their analysis by using information from the full spectrum of invariant mass or transverse mass, by including detailed estimates of systematic uncertainties, and by estimating the reach at 100 TeV\@.  The full NLO EW corrections from SUSY on $W$ and $Z/\gamma$ production have been computed by Refs.~\cite{Brensing:2007qm} and~\cite{Dittmaier:2009cr}, respectively.  We extend the above works by adopting a model-independent perspective which we define in more detail in \sec{EWrunning}.

The organization of this paper is as follows.  In \sec{EWrunning}, we review running EW couplings in the SM and in the presence of new particles.  We discuss the interplay of modified gauge coupling running and electroweak precision tests.    In \sec{DY}, we evaluate the cross sections for $Z^*/\gamma^*$ and $W^*$ production at various center of mass energies, and we discuss how the cross sections depend on the running of EW couplings.  We also review the status of theory predictions for Drell-Yan.    In \sec{uncertainties}, we discuss sources of uncertainty that will impact the precision with which $\alpha_{1,2}(Q)$ can be measured, including experimental uncertainties, theory uncertainties, and uncertainty from background subtraction.  We argue that percent-level measurements of $\alpha_{1,2}(Q)$ are possible, after accounting for sources of uncertainty.   In \sec{limits}, we present our main results.  We use existing LHC data to perform the first fits to running EW couplings above $\MZ$, and to provide model-independent limits on new states with EW quantum numbers.   We also present the reach at $\sqrt s = 14$ and 100~TeV\@.   \sec{conclusions} contains our conclusions.  We have also included several appendices of a more technical nature.  \app{beta} describes our notation and conventions for beta functions, \app{rescaling} provides an analytic understanding of how the neutral current Drell-Yan cross section depends on $\alpha_{1,2}(Q)$, \app{thy} provides further technical details on Drell-Yan theory predictions and uncertainties, \app{stats} describes the statistical procedure we use to set limits and estimate reach for measurements of running EW couplings, and \app{EWscale} discusses the sensitivity of the limits on new states to the scale dependence of the EW couplings.

\section{Running Electroweak Couplings}
\label{sec:EWrunning}
We focus on the inference of new particles based on their contribution to the evolution of couplings above their mass thresholds.  It is in this sense that we consider our approach as model-independent, as such a procedure is insensitive to how these states decay: our setup must therefore assume {\it only} that the dominant coupling of new states is to EW gauge bosons such that, for instance, four-fermion operators are not generated at tree-level below the new states' masses.  Under this assumption, our program requires no further input to constrain these states.  If, rather, new states generate contact operators between SM fermions at tree-level, their effect on the mass distributions  that we consider will appear as a power law rather than as a logarithmic one, allowing a discrimination of these scenarios through precise determination of event shapes.  Quantifying this precision comprises the bulk of Secs.~\ref{sec:DY} and  \ref{sec:uncertainties}, with results in Sec.~\ref{sec:limits} ultimately supporting the expectation that sufficiently clean channels can indeed be used to observe logarithmic modifications.

We work throughout to leading log order, in which case our setup is fully specified by new states' masses $M$ and contributions $\Delta b_{1,2}$ to the $U(1)$ and $SU(2)$ beta functions, i.e. we  assume a simplified model where deviations in weak coupling take a simple functional form 
\beq
\delta \alpha_{1,2} = \delta \alpha_{1,2}(M, \Delta b_1,\Delta b_2).
\eeq  
Beyond leading log order, the spin of the new EW states must also be included in order to fully specify the simplified model, due to the fact that the finite terms in vacuum polarization corrections arising from bosons in a loop differ from those arising from fermions. 

Here we first discuss details of the appearance of new EW states as seen through running couplings, and compare briefly to effects whereby precision measurements made {\it below} threshold may be sensitive to new states in an analogous way.   We employ a mass-independent subtraction scheme throughout, such that beta functions undergo discrete changes across mass thresholds; details of these conventions for encoding coupling evolution in beta functions are included in \app{beta}.

\subsection{Coupling Evolution in the SM and Beyond}
\label{subsec:UVrunning}
We focus first on the case of high energy indirect sensitivity to particles, at scales where they can contribute to running couplings. 
We will be interested in the constraining power of colliders operating above a particle's mass threshold, in the space of new contributions to EW beta functions, $\Delta b_{1,2}$.  As reviewed in the appendix, in the SM at scales above $ m_t$ the one-loop beta function values are given by $b_1 = 41/10$ and $b_2=-19/6$, with any new states contributing an amount proportional to the square of their charges.

If we assume that the pertinent cross sections used for  constraining evolution can be measured with percent-level accuracy,  sensitivity to differential deviations of a given $\alpha_i = g_i^2/4\pi$ will be of order $\delta \alpha/ \alpha \gtrsim 5 \times 10^{-3}$ because $\sigma \propto \alpha^2$.  In the cases studied below, e.g. $\df \sigma/\df \Mll$ for Drell-Yan dileptons at the LHC, the accuracy of measurement is indeed $\delta  \sigma/\sigma \approx 1\%$, so this seems a worthwhile figure of merit for the variation of the couplings themselves.

\begin{figure}[htb]
\begin{center}
\includegraphics[width=\textwidth]{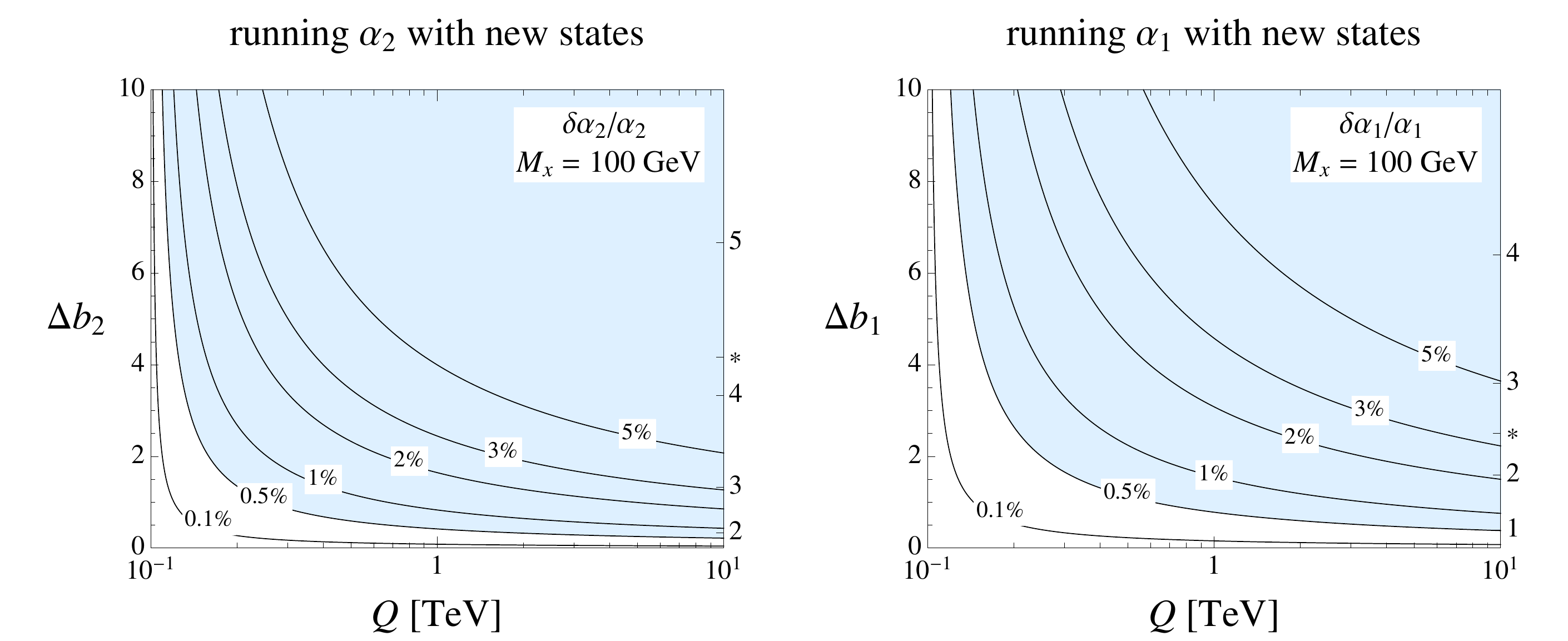} 
\caption{\small {Differential running of EW couplings, relative to SM, for varying renormalization scale, assuming new physics states enter at 100 GeV\@.  The shaded region indicates the space accessible provided percent-level accuracy is achieved in the measured cross sections.  The right axes indicate the corresponding representations for a given value of $\Delta b$ assuming the new states to be fermionic;  an asterisk in each case indicates the value of $\Delta b$ corresponding to the full matter content of the MSSM\@.
}}
\label{fig:bvsmu}
\end{center}
\end{figure}

\begin{figure}[htb]
\begin{center}
\includegraphics[width=\textwidth]{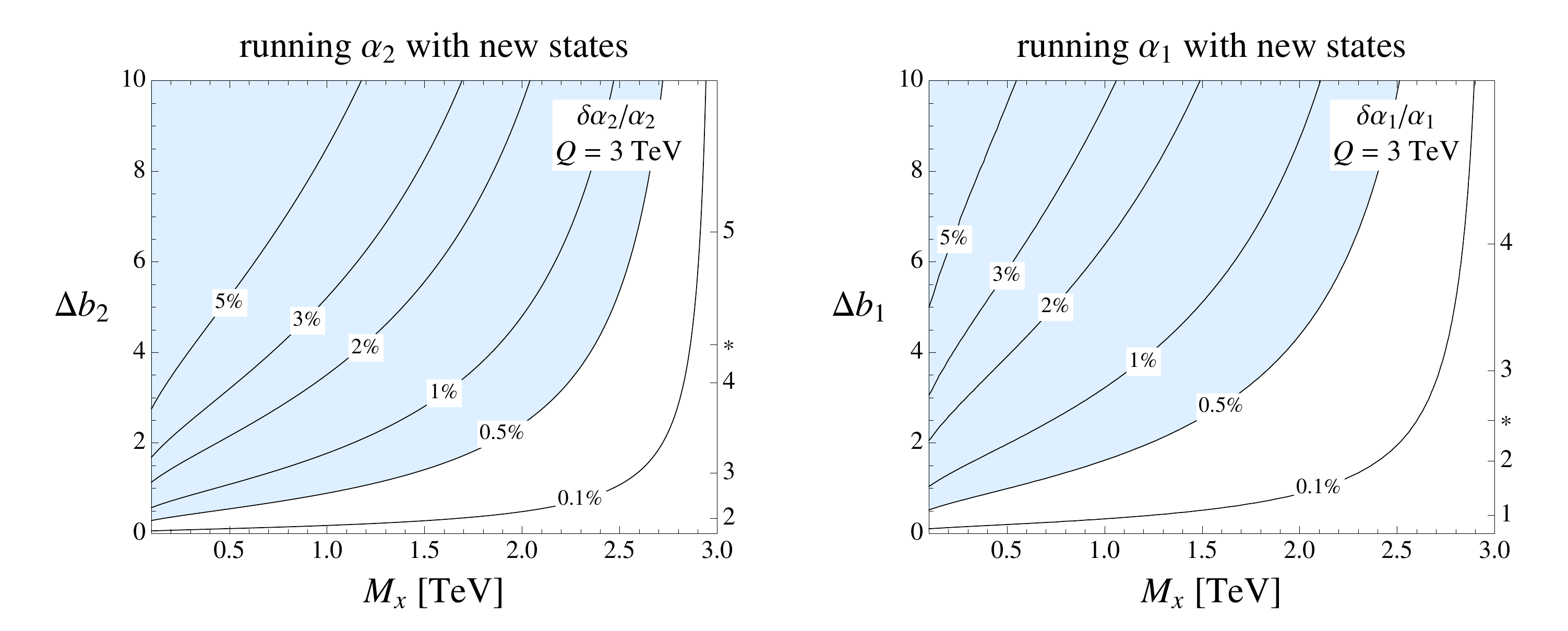} 
\caption{\small {Differential running of EW couplings for fixed renormalization scale $Q = 3$~TeV, allowing the mass of new states, $M_x$, to vary.  Shading and labeling are as in \fig{bvsmu}.}}
\label{fig:bvsM}
\end{center}
\end{figure}

For the sake of illustration, in \fig{bvsmu} we assume generic new physics entering at a scale $M_X = 100\, {\rm GeV}$ and show how it affects the gauge couplings as a function of renormalization scale $Q$ as its contribution to the beta functions is varied.  
Alternatively we can fix a renormalization scale and plot the same effect as a function of the mass of the new states.  We show the latter in \fig{bvsM}, taking $Q =3\, {\rm TeV}$ as the approximate scale below which precision studies are possible with a hadron collider operating at center of mass energy $\sqrt s = 100 \, {\rm TeV}$.  

\subsection{Scale Dependence of Processes Below Threshold}
\label{subsec:IRrunning}
Our renormalization scheme requires heavy particles to be integrated out manually, their impact on the low energy physics entering only through higher dimension operators.  In the case of running EW couplings, there is a natural choice of such operators one might hope to constrain with low energy precision measurements that could allow comparison with the indirect tests at high energy that we advocate above.  In particular, the parameters $W$ and $Y$ of~\cite{Barbieri:2004qk} must be counted among the minimal set for electroweak precision constraints~\cite{Cacciapaglia:2006pk}.  These particular parameters are defined as
\beq
W = \frac{g^2 \MW^2}{2} \Pi''_{W_3 W_3}(0),
\quad 
Y = \frac{g'^2 \MW^2}{2} \Pi''_{BB}(0), 
\eeq
with $\Pi_{VV'}(q^2)$ specifying the two-point function among gauge fields $V, \, V'$.\footnote{These expressions are equivalent to treating $W$ and $Y$ defined as proportional to the coefficients of dimension six operators $(D_\rho W_{\mu \nu}^a)^2$ and $(\partial_\rho B_{\mu \nu})^2$, respectively.}
Explicit calculation of these functions, in the case where  new physics is assumed to enter only at loop level through gauge couplings, gives
\beq
\Pi''_{VV}(0) = \left(\frac{2}{3}\sum_{f,\, r} + \; \frac{1}{3} \sum_{s,\, r}  \right) \frac{T_r}{40 \pi^2 M_x^2} 
\eeq
for fermions and scalars respectively appearing in representations $r$.  This allows us to directly compare the electroweak precision parameters to the space of beta functions:
\beq
\label{eq:WtoBeta}
W,\, Y  = \frac{\alpha_{2,1}}{20 \pi} \frac{\MW^2}{M_x^2} \times \Delta b_{2,1}.
\eeq
These parameters give the leading constraints for states on which custodial isospin and EW symmetry are preserved. Following~\cite{Barbieri:2004qk}, 
\begin{table}[htb]
\begin{center}
\renewcommand{\arraystretch}{1.2}
\begin{tabular}{| c || c | c | c || c | c |}
\hline
  &  $\Delta \MW \ {\rm (GeV)}$ & $ \Delta \Gamma_Z \ {\rm (GeV)}$ & $\Delta A_{\rm LR}$ & $ \Delta W$ & $\Delta Y$   \\
\hline
\ \ EWPT+LEP2~\cite{LEP,Altarelli:1990zd, Baak:2012kk} \ \ & \ \  $3.4\cdot 10^{-2}$ \ \ & \ \ $ 2.3\cdot 10^{-3}$ \ \ & \ \ $ 2.1\cdot 10^{-3}$ \ \ & \ \ $8\cdot 10^{-4}$  \ \ &  \ \  $1.2\cdot 10^{-3}$ \ \ \\
\ \ ILC (GigaZ)~\cite{ILC} \ \ & $\ \ 6\cdot 10^{-3}$ \ \ & $\ \ 8 \cdot 10^{-4} $ \ \ & $  10^{-4} $ & $\ \ 3 \cdot 10^{-4} $ \ \  & $\ \ 3 \cdot 10^{-4} $ \ \ \\
\ \ TLEP (TeraZ)~\cite{TLEP} \ \ & \ \ $  5 \cdot 10^{-4} \ {\rm (sys)}$ \ \ & \ \ $ 10^{-4}\ {\rm (sys)}$  \ \ & $\ \ 1.5 \cdot 10^{-5} $\ \ & $\ \ 7 \cdot 10^{-5} $ \ \ & $\ \ 1 \cdot 10^{-4} $ \ \ \\
\hline
\end{tabular}
\caption{\small {Current and projected uncertainties on observables and derived quantities, from electroweak precision tests (EWPT) and LEP2 and for future lepton colliders. The dominant uncertainties are statistical unless noted otherwise.}}
\label{tab:Pdata}
\end{center}
\end{table} 
we show in \tab{Pdata}  how  $W$ and $Y$  respond to reduced uncertainties on $Z$-pole   observables to which they are most sensitive.  In particular we find that a reduction of these observables' uncertainties, with an increased correlation between them, amounts to a typical improvement in electroweak precision parameters of order $\sim3-4$ at the ILC and $\sim8-12$ at TLEP; projections for the $S$ and $T$ parameters reported elsewhere display this behavior as well (see {\it e.g.}~\cite{Baak:2014ora,Henning:2014gca}).  We will show in \sec{limits} how these results compare to constraints arising from running couplings measured at hadron colliders.

\section{Measuring Running Electroweak Couplings with Drell-Yan}
\label{sec:DY}

\begin{figure}[htb]
\begin{center}
\includegraphics[height=2.3cm]{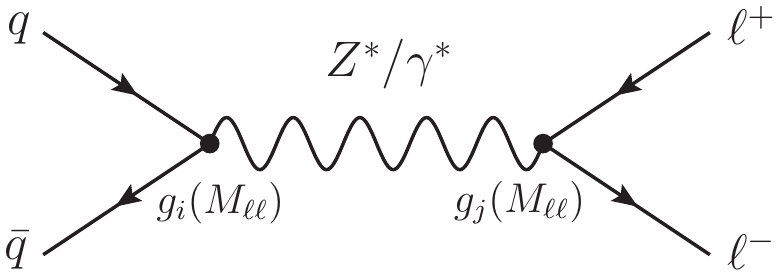} \qquad \qquad
\includegraphics[height=2.3cm]{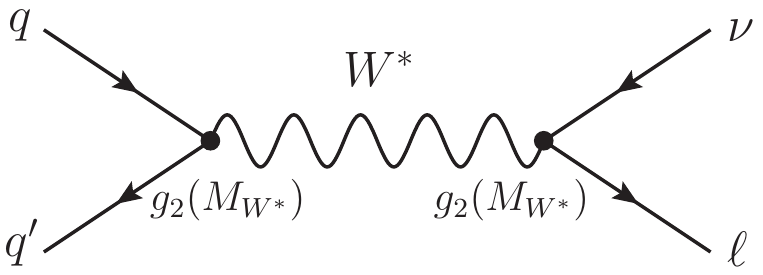}
\caption{\small {Neutral and charged current Drell-Yan processes used for measurement and constraints of weak couplings, $\alpha_{1,2}$, at hadron colliders.  The neutral process probes both couplings; the charged process probes $\alpha_2$ alone.}}
\label{fig:DYdiagram}
\end{center}
\end{figure}

As discussed in \sec{EWrunning}, measuring the scale dependence of electroweak couplings above the weak scale requires a high precision EW process.  At a hadron collider the Drell-Yan processes, $\Zproc$ and $\Wproc$, illustrated in \fig{DYdiagram},
are  ideal candidates: the event rate is substantial even out to large energy scales, the leptonic final state is experimentally clean, and the cross section is proportional to $\alpha_{1,2}^2$ which allows for a simple measurement of the coupling.  Current measurements at the LHC show that the cross section can be measured with near percent-level systematic uncertainties over a wide range of energies above the weak scale, and the theoretical uncertainties on the cross section are similarly small.  With similar performance at future high-energy colliders, there is a good opportunity to expand the reach of the measurement well into the TeV range.

\begin{table}[t!]
\centering
\renewcommand{\arraystretch}{1.2}
\begin{tabular}{|c|c|}
\hline
cuts \, ($l=e,\mu$) & luminosity ($\mathcal{L}^{\sqrt s}$) \\ 
\hline
\ $p_T^l > 25$ \qquad $|\eta^l|< 2.5$ \ & $\mathcal{L}^{8} =20$~fb$^{-1}$ \\
$M_{ll}^{Z^*} > 125$~GeV  &\  $\mathcal{L}^{14} = 300(0)$~fb$^{-1}$  \ \\
$M_T^{W^*} > 125$~GeV  &  $\mathcal{L}^{100} = 3000$~fb$^{-1}$ \\
\hline
\end{tabular}
\caption{\small {Kinematic cuts and luminosities used for our numerical studies.}}
\label{tab:cuts}
\end{table}

\subsection{Neutral Current Drell-Yan}
\label{subsec:Zgamma}

For the process $\Zproc$ with $\ell = e$ or $\mu$, the leptonic final state can be identified with high precision and the EW couplings probe the offshellness of the vector boson in the quark annihilation and lepton production.  A natural choice for the renormalization scale, which we make here, is the invariant mass of the dilepton pair, $\Mll$.  Therefore the differential cross section, making the LO coupling dependence explicit,
\begin{equation} \label{eq:sigmaZgamma}
\frac{\df\sigma}{\df \Mll} (\Zproc) \equiv \frac{\df\sigma^{Z/\gamma}}{\df \Mll} \bigl(\alpha_{1,2} ( \Mll )\bigr)
\end{equation}
provides direct sensitivity to the electroweak couplings at $\Mll$.

\begin{figure}[b!]
\begin{center}
\includegraphics[height=5.25cm]{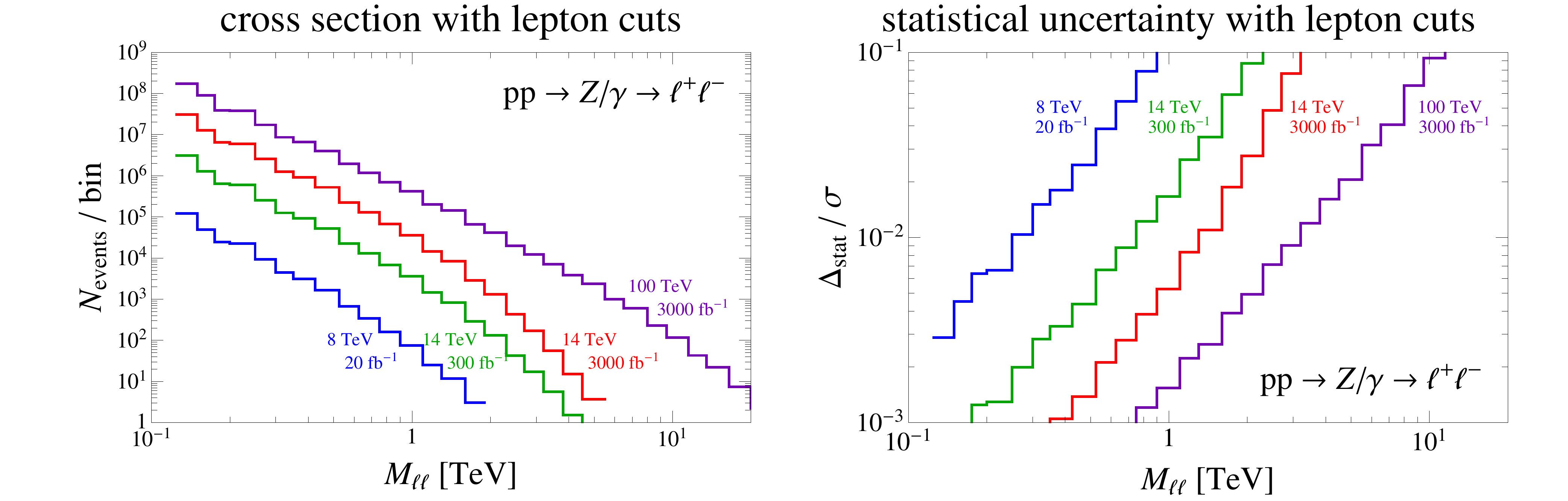}
\caption{\small{On the left, the expected number of events per bin for $\Zproc$ with nominal acceptance cuts on the final state leptons, for various center of mass energies and benchmark luminosities.  On the right, the corresponding statistical uncertainty in each bin for each energy/luminosity combination.  The NNLO QCD cross section is used.}}
\label{fig:NevDY}
\end{center}
\end{figure}

The $\Zproc$ cross section is measured with acceptance cuts designed to select only those events with a high resolution dileptonic final state.  Typically this involves a minimum $p_T$ cut and a pseudorapidity cut on each lepton to select central, high $p_T$ leptons that can be cleanly identified.  For our studies we will define simple acceptance cuts of $p_{T\ell} > 25 \text{ GeV} \, \lvert \eta_\ell \rvert < 2.5$, which we summarize in Table~\ref{tab:cuts}.
The efficiency to pass these cuts is $\ord{50\%}$ around the weak scale and goes to 1 at large $\Mll$.  In our studies of neutral current Drell-Yan, we consider final states in $e^+e^-$ and $\mu^+\mu^-$, with invariant mass $\Mll>125~\GeV$. In \fig{NevDY}(left) we show the invariant mass distributions in the neutral Drell-Yan channel for the different center-of-mass energies/luminosity scenarios we will investigate, namely, $\sqrt{s}=8~\TeV$ with 20~$\text{fb}^{-1}$ of integrated luminosity, 14 TeV with 300~$\text{fb}^{-1}$ and 3000~$\text{fb}^{-1}$, and 100 TeV with 3000~$\text{fb}^{-1}$. In \fig{NevDY}(right) we show the size of the statistical uncertainties relative to the cross section, from which we can infer the mass scale above which sensitivity is lost to percent-level effects in the Drell-Yan cross section.

The differential cross section for this process is sensitive\footnote{A more exclusive cross section, such as one also differential in the positively charged lepton rapidity, may be used to differentiate between the $\alpha_1$ and $\alpha_2$ dependence.} to both $\alpha_1$ and $\alpha_2$.  
In order to probe this two-dimensional space in an efficient way, it is convenient to consider rescaling 
the weak mixing angle (denoted $\hat s_W = \sin \hat \theta_W$)\footnote{The notation used here is chosen to coincide with that commonly used in $\overline {\rm MS}$, where 
$$
\sin \hat \theta_W (Q) \equiv \frac{g'(Q)}{\sqrt{ g^2(Q) + g'^2(Q) }}.
$$
This highlights the fact that variations in $\hat s_W$ will ultimately be interpreted as  constraining  the values of  running couplings within this scheme.}, 
along with the combination
\beq
\label{eq:aalpha}
a = \frac{\alpha_2}{\hat c_W^2} = \frac{3}{5} \frac{\alpha_1}{\hat s_W^2} = \frac{\alpha_{\rm EM}}{\hat s_W^2 \hat c_W^2}.
\eeq
All parameters  here implicitly dependent on the renormalization scale.
This leads to a simple functional form of rescaling for the partonic cross section:
\beq
\df  \sigma^{Z/\gamma}   = a^2 f(\hat s_W^2, \Mll;Q) \times \df  \sigma_0^{Z/\gamma},
\eeq
where at tree level we have
\beq
\label{eq:fr}
f(\hat s_W^2,\Mll;Q) = c_0 + c_1 \hat s_W^2 + c_2 \hat s_W^4 + c_3 \hat s_W^6 + c_4 \hat s_W^8.
\eeq
All dependence on $\Mll$ is captured by the coefficients $c_i$ and the proportionality constant $\df  \sigma_0^{Z/\gamma}$,  expressions for which we collect in \app{rescaling}.

At scales $Q \gg \MZ$, the rescaling function $f(\hat s_W^2, \Mll; Q)$ reduces to a quadratic function of $\hat s_W^2$.  This is understood  from the fact that the $Z$ and $\gamma$ are both effectively massless in this limit, allowing use of a basis where all contributions to the cross section scale simply as $\alpha_{1,2}^2$ or $\alpha_1 \alpha_2$.  Terms at higher order in $\hat s_W^2$ become important near the $Z$ pole through mixing angles, though their contribution to the cross section reaches (sub)percent levels already at  $Q \gtrsim 300\, {\rm GeV}$.  Even in full generality, however, the upshot remains that complete knowledge of the leading order rescaling in the space of $(\alpha_1, \alpha_2)$ for a given $\Mll$ is provided simply by fitting the coefficients $c_i$.  Working at tree-level in the weak couplings, we fit the coefficients of the rescaling function for each mass bin from tree-level matrix elements by appropriately varying input parameters within {\tt MadGraph5} v1.3.30~\cite{Alwall:2011uj}.

\subsection{Charged Current Drell-Yan}
\label{subsec:W}

In the process $\Wproc$, the final state charged lepton is most cleanly identified if it is an electron or muon, while the neutrino creates missing transverse energy (MET) in the reconstructed event.  Therefore the dilepton pair invariant mass may not be reconstructed, but the transverse mass, $\MT$, is a suitable variable to probe the large momentum exchange regime of the cross section.  For the massless dilepton final state, $\MT$ is defined as
\begin{equation}
\MT^2 = 2 ( p_{T\ell} \, p_{T\nu} - \vec{p}_{T\ell} \cdot \vec{p}_{T\nu} ) \,,
\end{equation}
where $\vec{p}_{T\nu} = \slash{\!\!\!\!\vec{E}}_T$ is the MET in the event.  The distribution of $\MT$ has an endpoint at $\Mlnu$, with $\MT \le \Mlnu$, but peaks near the endpoint at values close to $\Mlnu$.  Therefore the differential cross sections, which at LO depend only on $\alpha_2$,
\begin{align} \label{eq:sigmaW}
\frac{\df\sigma}{\df \MT} (\Wproc) &\equiv \int_{\MT}^{\infty} \df \Mlnu \frac{\df\sigma^{W^\pm}}{\df \MT \df \Mlnu} \bigl(\alpha_2 (\Mlnu) \bigr) \nonumber \\
&= \frac{\df\sigma^{W^\pm}}{\df \MT}
\end{align}
provide direct sensitivity to the electroweak couplings at a scale which is \emph{at least} $\MT$.  We note also that because of this simple $\alpha$-dependence, rescaling the cross section within a given bin of $\Mlnu$ according to varying beta functions faces none of the subtleties present in the neutral current case; the only distinction is that the rescaling effects must be mapped from $\Mlnu$ to $\MT$.

\begin{figure}[htb]
\begin{center}
\includegraphics[height=5.25cm]{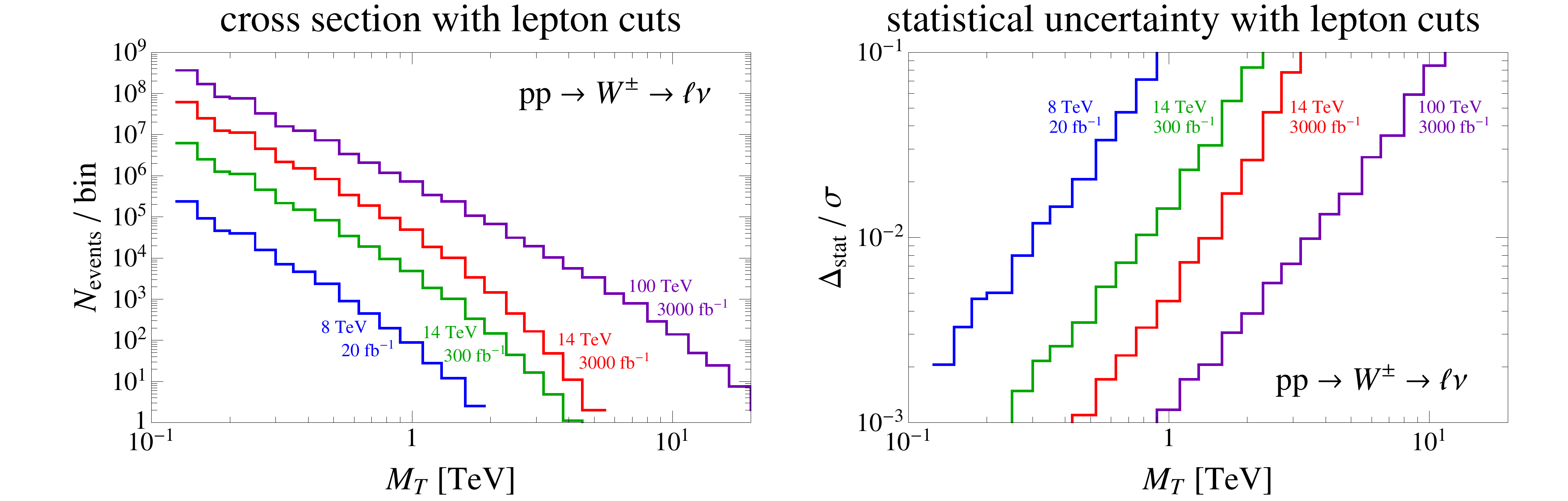}
\caption{\small{On the left, the expected number of events per bin for $\Wproc$ with nominal acceptance cuts on the final state charged lepton, for various center of mass energies and benchmark luminosities.  On the right, the corresponding statistical uncertainty in each bin for each energy/luminosity combination.  The NNLO QCD cross section is used.}}
\label{fig:NevWtot}
\end{center}
\end{figure}

The $\Wproc$ cross sections are measured in final states with a charged lepton ($e^\pm$ or $\mu^\pm$) and missing transverse energy ($\MET$).  For our studies we will use the same acceptance cuts on the final state charged lepton as in the $Z/\gamma$ process, namely, $
p_{T\ell} > 25~\GeV$ and $|\eta_\ell| < 2.5$, and consider the transverse mass spectrum $M_T>125~\GeV$,  as shown in Table~\ref{tab:cuts}. We will also assume that the $\ell^++\MET$ and $\ell^-+\MET$ final states will be measured separately. Our requirement of large transverse mass above $M_W$ provides an implicit requirement on minimum value of $\MET$, so we do not place further cuts on the missing transverse energy.  The efficiency to pass the acceptance cuts is $\ord{70\%}$ and quickly saturates to 1 as $\MT$ increases.  In \fig{NevWtot} we show the transverse mass distributions in the charged Drell-Yan channel for the same scenarios as in neutral Drell-Yan (namely, 8~TeV with 20~$\text{fb}^{-1}$, 14 TeV with 300~$\text{fb}^{-1}$ and 3000~$\text{fb}^{-1}$, and 100 TeV with 3000~$\text{fb}^{-1}$), as well as the statistical uncertainties relative to the cross section.

\subsection{Theory Predictions}
\label{subsec:theory}

The Drell-Yan processes are among the most thoroughly-studied and well-understood at hadron colliders.  The most important theory ingredients in accurately calculating the cross section are fixed-order QCD and EW corrections and precise determinations of the parton distribution functions (PDFs).  The QCD and EW corrections have been determined at NNLO (see Refs.~\cite{Hamberg:1990np,Anastasiou:2003yy,Anastasiou:2003ds,Melnikov:2006kv,Catani:2009sm,Gavin:2010az,Li:2012wna}) and NLO (see Refs.~\cite{Berends:1984xv,Baur:1997wa,Baur:2001ze,CarloniCalame:2007cd,Arbuzov:2007db,Dittmaier:2009cr} and Refs.~\cite{Wackeroth:1996hz,Baur:1998kt,Dittmaier:2001ay,Baur:2004ig,Arbuzov:2005dd,CarloniCalame:2006zq}) respectively, and PDFs have been fit through NNLO\@.  Although the QCD corrections have been computed to higher order, and have larger scale uncertainties, in the high-mass regime of the cross section the EW corrections can make numerically important contributions from EW Sudakov logarithms and photon-initiated production of $\ell^+\ell^-$.

\begin{figure}[htb]
\begin{center}
\includegraphics[height=5.25cm]{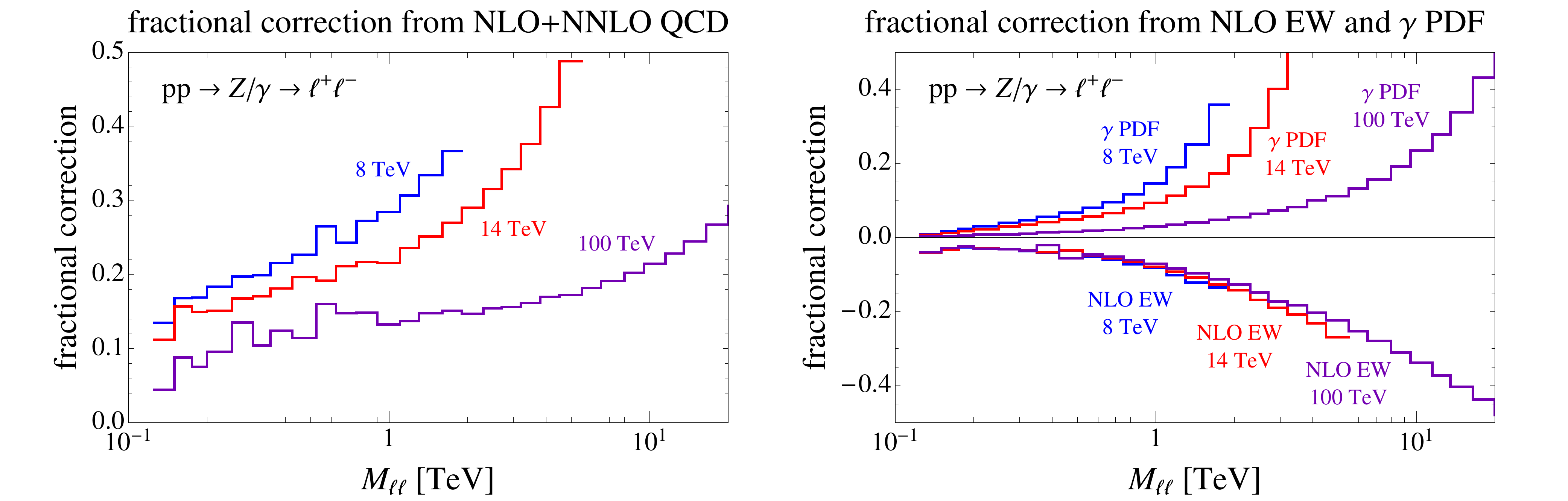}
\caption{\small {The fractional corrections relative to the LO cross section of the NLO + NNLO QCD (left) and NLO EW (right) contributions to $pp \to \ell^+ \ell^-$.  In the case of EW corrections, we separate the contributions into EW corrections to $\Zproc$ and the contribution of photon-initiated direction production of $\ell^+ \ell^-$, which requires a photon PDF\@.  The fluctuations at low invariant mass are statistical in nature.}}
\label{fig:ZprodKfac}
\end{center}
\end{figure}

To study the NNLO QCD and NLO EW corrections, we use the \texttt{DYNNLO} and \texttt{FEWZ} generators~\cite{Catani:2007vq,Catani:2009sm,Melnikov:2006kv,Gavin:2010az,Gavin:2012sy,Li:2012wna}, both of which can evaluate fully exclusive Drell-Yan cross sections through NNLO in QCD\@.  \texttt{FEWZ} additionally includes the NLO EW corrections for neutral current Drell-Yan, specifically the contributions from virtual $W$s and $Z$s as well as virtual and real QED corrections.  These corrections can be substantial in the high mass regime of the cross section, and understanding these corrections is important for having control over the theoretical uncertainties.  In \fig{ZprodKfac}, we plot the higher order QCD and EW fractional corrections to the LO neutral current cross section.  To evaluate these contributions, the NNPDF 2.3 NNLO PDF sets with and without QED corrections were used, with $\as(\MZ) = 0.119$~\cite{Ball:2012cx,Ball:2013hta}.

In the case of QCD, the dominant contributions are coming from the NLO corrections. The full NNLO cross section is well controlled with uncertainties in the $\ord{1-2\%}$ range, except at the largest invariant masses where statistical uncertainties dominate.  As we will see, this uncertainty is of the same order as the PDF uncertainty, and the two comprise the largest theoretical uncertainties on the cross section.

The EW corrections are more subtle.  It is well known that electroweak Sudakov logarithms exist due to exclusive constraints on the EW final state (see, e.g., Refs.~\cite{Baur:2006sn,Bell:2010gi,Becher:2013zua}, relevant to the Drell-Yan process), and exist even in inclusive cross sections due to the fact that the initial states that are not electroweak singlets~\cite{Ciafaloni:2000df,Ciafaloni:2000rp,Ciafaloni:2000gm,Ciafaloni:2001vt}.  The form of these logarithms, and their resummation, can differ process-to-process, and requires careful treatment of the real electroweak radiation to ensure proper cancellation with numerically large contributions from virtual corrections (see, e.g.,~\cite{Chiu:2007yn,Chiu:2007dg,Chiu:2008vv,Chiu:2009mg,Chiu:2009ft} for a general discussion of EW Sudakov logarithms in virtual corrections, or Ref.~\cite{Manohar:2014vxa} for a recent study on a set of specific processes).  In the cross section $\df\sigma^{Z/\gamma} / \df \Mll$, these contributions manifest as logarithms of the form $\ln \Mll^2 / \MZ^2$, and in \fig{ZprodKfac} these logarithms are the source of the increasing NLO EW corrections at larger invariant masses.  

Real weak radiation mixes the charged and neutral current Drell-Yan channels.  For example, $\Zproc$ production with a subsequent $W$ emission from the final state leptons will convert that charged lepton into a neutrino, leading to a $\ell\nu W$ final state, and a hadronic $W$ decay will give a charged-current-like final state.  Exclusive cuts on the number of leptons or cuts on missing energy will remove a significant fraction of events that contain real weak radiation.  This implies that real radiation will only partially cancel the virtual corrections, and so large logarithmic effects can persist in measured cross sections.  This was nicely studied in Ref.~\cite{Baur:2006sn} for a variety of processes that include Drell-Yan, and it was found that real weak radiation introduces only an $\ord{20\%}$ cancellation of the virtual weak and NLO electromagnetic corrections for a certain set of reasonable cuts at a 14 TeV collider.

In principle, this presents a challenge for precise predictions of the Drell-Yan cross sections at high invariant mass.  To set robust limits and make a sensitive measurement, the EW uncertainties must be under good control, especially at intermediate masses where the NLO corrections are non-negligible but the statistical uncertainties do not yet dominate.  Resummation of the EW Sudakov logarithms will bring perturbative stability to the cross section, and in Refs.~\cite{Chiu:2008vv,Chiu:2009mg} it was suggested that this resummation along with QCD corrections can achieve a scale uncertainty within 1\%.  Hence, while these corrections are numerically important, resummation of these effects will give a well-controlled theoretical prediction for the EW corrections, and one expects small uncertainties on the result.  Such resummation is beyond the scope of the work here, and to our knowledge has not been preformed for the cross sections $\df\sigma^{Z/\gamma} / \df \Mll$ and $\df\sigma^{W} / \df \MT$.

In addition to resummation of the EW Sudakov logarithms, constructing a careful inclusive cross section can maximize the potential for real emissions to cancel the Sudakov logarithms from virtual corrections.  Non-electroweak singlet initial states generate Bloch-Nordsieck violations~\cite{Ciafaloni:2000df}, which subsequently lead to EW Sudakov logarithms in the cross section.  Additionally, though, separating the charged and neutral current channels limits the real corrections to each channel.  Therefore, we can combine channels by treating neutrinos and leptons equivalently, using the observable
\begin{equation}
\MT^{\rm EW} \equiv \sqrt{ 2 ( p_{T,L_1} \, p_{T,L_2} - \vec{p}_{T,L_1} \cdot \vec{p}_{T,L_2} ) } \,,
\end{equation}
where $L_1$ and $L_2$ are chosen from the set $\{ \text{MET}, \ell_1, \ldots, \ell_n \}$, which is the set of missing transverse energy and all charged leptons (passing acceptance cuts), such that $\MT^{\rm EW}$ is maximized.  This observable includes contributions from both neutral and charged current Drell-Yan production, and at leading order has the value
\begin{equation}
\frac{\df\sigma}{\df \MT^{\rm EW}} = \frac{\df\sigma^{Z/\gamma}}{\df \MT^{\ell\ell}} + \frac{\df\sigma^{W^\pm}}{\df \MT^{\ell\nu}} \,.
\end{equation}
Furthermore, this observable is inclusive over weak boson emission (that does not push the final state leptons below acceptance cuts, which is generally not a concern for large $\MT^{\rm EW}$ values), meaning that real and virtual corrections will cancel substantially more than in the separate charged and neutral current channels.  Residual EW Sudakov logarithms will be largely limited to Bloch-Nordsieck violating terms, and the electroweak corrections will be under better control.  Finally, $\MT^{\rm EW}$ is a simple way to combine the power of the Drell-Yan channels in constraining the running of $\alpha_{1,2}$.

Part of the NLO EW corrections open new partonic channels for the cross section, those coming from photon-initiated processes.  In the neutral current case, this channel is $\gamma \gamma \to \ell^+ \ell^-$, and requires a photon PDF for the proton.  Although theoretical considerations and fits to HERA data suggest the photon distribution is small~\cite{Martin:2004dh,Martin:2014nqa}, fits that largely depend on ATLAS neutral current Drell-Yan data suggest that it may have a large contribution in the high dilepton invariant mass region of the cross section, with large uncertainties~\cite{Ball:2013hta}.  Most importantly for our discussion, because the photon PDF is difficult to measure, and the available data to do so is limited, the uncertainty on the photon PDF extracted from LHC data is substantial, $\ord{100\%}$ of its contribution.  The photon PDF may be better measured via more exclusive Drell-Yan observables sensitive to its contribution, such as the $p_T$ spectrum of single charged leptons~\cite{Dittmaier:2009cr,Boughezal:2013cwa}.  Furthermore, increased data from the LHC will significantly constrain future determinations of the PDF, so that one expects the uncertainty to correspondingly decrease in the future.  In our analysis, we will assume that the photon PDF will be measured sufficiently to reduce the uncertainty below those of the leading theoretical uncertainties on the cross section (see \subsec{unknowns} for further discussion).

Accurate theory predictions of the cross sections $\df\sigma^{Z/\gamma} / \df \Mll$ and $\df\sigma^{W^\pm} / \df \MT$ are needed to make precise measurements of the electroweak coupling at a given scale.  However, the relevant theory ingredients are determined by those whose contributions exceed or are competitive with the dominant theoretical uncertainties.  Furthermore, for the purpose of estimating the reach of future measurements of electroweak couplings, only the uncertainties on the theoretical predictions are relevant, as they (along with experimental uncertainties) control the reach of the measurement.  The QCD scale and PDF uncertainties are the most significant theory uncertainties in the Drell-Yan cross sections, and accurate evaluation of them is crucial.  In examining the theoretical predictions for the Drell-Yan cross sections, our goal is not to arrive at the most accurate available theory predictions but instead to quantify the extent that various contributions to the cross section are under control (and those that are not), so that the theoretical uncertainties can be robustly quantified.

\section{Uncertainties in Measuring Electroweak Couplings}
\label{sec:uncertainties}

\subsection{Theory Uncertainties}
\label{subsec:theoryunc}

The dominant theoretical uncertainties on the Drell-Yan cross sections are the QCD scale and PDF uncertainties.  Although the EW corrections are substantial at large invariant mass, as discussed in \subsec{theory} we will assume that resummation of the electroweak Sudakov logarithms in the Drell-Yan cross sections differential in $\Mll$ or $\MT$ have subleading uncertainties.  Additionally, we assume that the photon PDF will be sufficiently measured to constrain its behavior at large invariant masses and reduce uncertainties below the leading ones.  These issues are further discussed in \subsec{unknowns}.

The QCD scale and PDF uncertainties are calculated by running the fixed order NNLO generators \texttt{DYNNLO} and \texttt{FEWZ}.  The QCD scale uncertainties are evaluated using the central renormalization and factorization scale choices
\begin{equation}
\mu^{\rm central}_R = \mu^{\rm central}_F = \Mll (Z/\gamma) \text{ or } \Mlnu (W^\pm) \,,
\end{equation}
and varying these scales collectively up and down by a factor of two, $\mu_{R,F}^{\rm up} = 2 \mu_{R,F}^{\rm central}$ and $\mu_{R,F}^{\rm down} = (1/2) \mu_{R,F}^{\rm central}$.  The maximum deviation of these variations sets the scale uncertainty,
\begin{equation}
\Delta_{\rm scale} = \max\bigl( \lvert \sigma(\mu_{F,R}^{\rm up}) - \sigma(\mu_{F,R}^{\rm central}) \rvert, \lvert \sigma(\mu_{F,R}^{\rm down}) - \sigma(\mu_{F,R}^{\rm central}) \rvert \bigr) \,.
\end{equation}

We use the NNPDF 2.3 NNLO PDF set, with $\as(\MZ) = 0.119$.  The NNPDF sets have members which form a statistical ensemble from which PDF uncertainties can be evaluated by taking the standard deviation of the cross section values given by those members.  We do not include $\as$ uncertainties in our estimation, as they will be subdominant to the other uncertainties.  In \app{thy} we give further details on the evaluation and features of the QCD scale and PDF uncertainties, in addition to the scale and PDF uncertainty correlations between bins of $\Mll$ or $\MT$.

\figs{ATLASunc}{ErrDy} include the QCD scale and PDF uncertainties as a function of $\Mll$ for $\Zproc$ production, and \fig{ErrW} includes the same uncertainties as a function of $\MT$ for $\Wproc$ production.

\subsection{Experimental Uncertainties and Backgrounds to Drell-Yan}
\label{subsec:experimentunc}

The experimental uncertainties to the measurement of Drell-Yan distributions come from experimental systematics, as well as the contamination of backgrounds to Drell-Yan final states. The dominant backgrounds to $Z^*/\gamma^*$ and $W^*$ Drell-Yan are QCD multijets in which one or more jets are misidentified as a lepton ($e^\pm$/$\mu^\pm$), as well as $\tau^+ \tau^-$, $t\bar{t}$ and diboson decaying leptonically. The reducible multijet backgrounds contribute to both the low and high invariant mass regions and require a data-driven estimation of fake rates. At high invariant mass, $t\bar{t}$ and diboson are the dominant irreducible backgrounds. Experimental systematics affect both the measurement of Drell-Yan processes as well as the estimation and subtraction of those backgrounds, and include uncertainties in: acceptance and efficiency in the event selection, energy scales, detector resolution, corrections for FSR and bin migration, fake rates, MC modeling, luminosity, as well as statistical uncertainties both in data and MC simulation. While systematic uncertainties are known for the LHC at 7~and~8~TeV, and can be realistically projected at 14~TeV, they are undetermined for the next generation of hadron colliders such as a future collider with $\sqrt{s}=100~\TeV$.

\begin{figure}[t]
\begin{center}
\includegraphics[height=6.4cm]{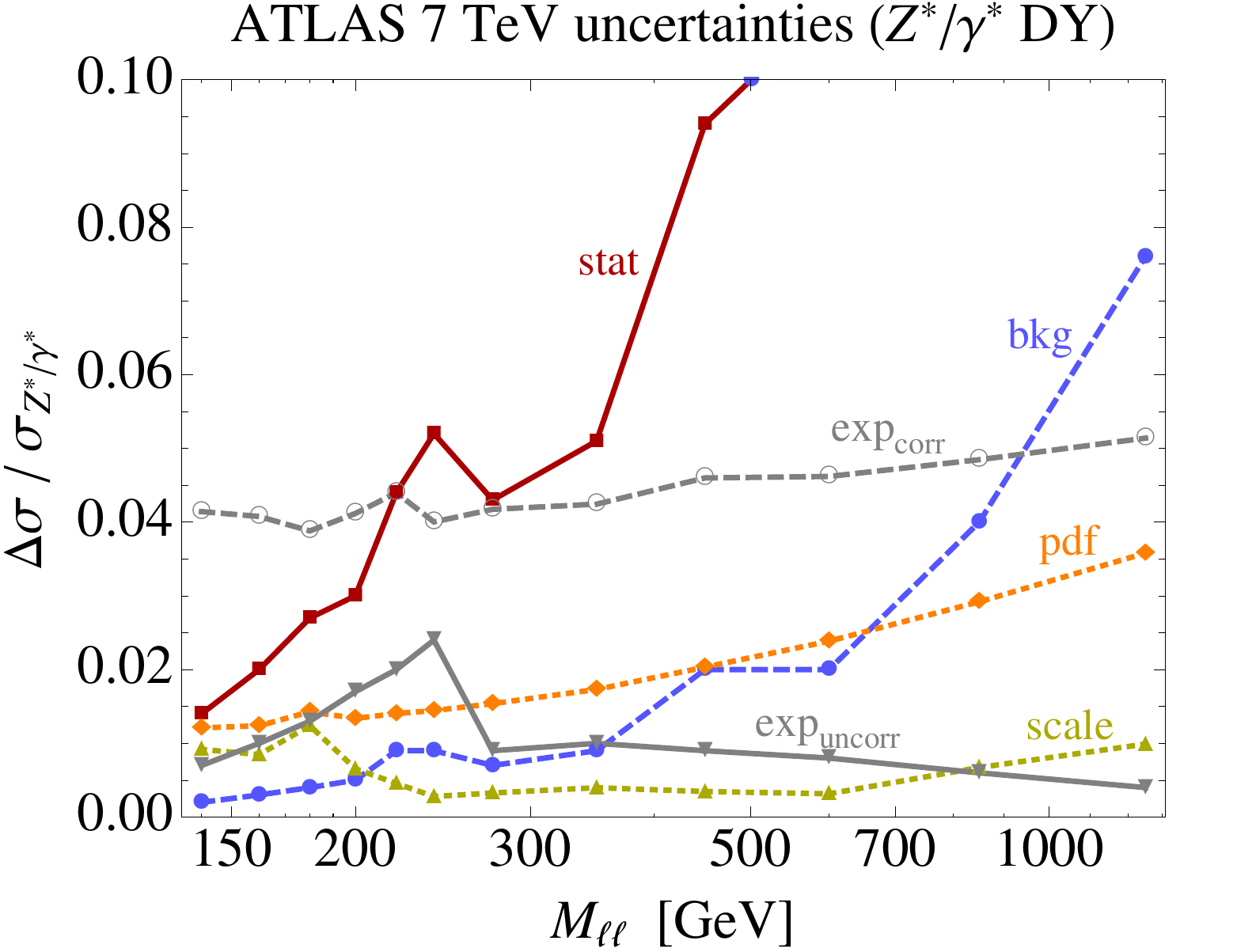}
\caption{\small {Uncertainties in 7 TeV neutral Drell-Yan measurements. The experimental uncertainties shown are from the ATLAS DY analysis~\cite{Aad:2013iua}, namely, statistical uncertainties (red), background subtraction uncertainties (blue), as well as uncorrelated (solid gray) and correlated (dashed gray) systematics. The theoretical uncertainties from PDF (orange) and scale (yellow) are estimated at fixed order NNLO with NNPDF2.3.}
\label{fig:ATLASunc}
}
\end{center}
\end{figure}

ATLAS and CMS have estimated their uncertainties both at 7 TeV~\cite{Aad:2013iua,Chatrchyan:2013tia} and 8 TeV~\cite{CMS:2014hga}, and found that backgrounds contribute 1 -- 5\% uncorrelated uncertainties in the invariant mass bins below 1 TeV and are subdominant to the statistical uncertainties. In addition, experimental systematics amount to $\mathcal{O}$(1\%) uncorrelated uncertainties across bins, and $\mathcal{O}$(5\%) correlated uncertainties. CMS has made publicly available all sources of uncertainties and their full correlation matrix for their 7~TeV Drell-Yan measurements~\cite{CMShepdata}, while ATLAS has provided a breakdown of the uncertainties in each bin modeled approximately as either fully uncorrelated with other bins or fully correlated across all bins~\cite{ATLAShepdata}. \fig{ATLASunc} displays the ATLAS 7~TeV breakdown of uncertainties for each invariant mass bin.

In conjunction with theoretical uncertainties previously discussed, having background and experimental systematics under control is essential in obtaining sensitivity to the running of the electroweak couplings in Drell-Yan measurements. Is it beyond the scope of this work to attempt to model all sources of experimental uncertainties, some of which depend on undetermined collider specifications and technology. In our sensitivity projections for 8,~14~and~100~TeV, we take the Drell-Yan systematic uncertainties at 7~TeV as a proxy for the order of magnitude of experimental systematics that we expect to be present in future measurements. In particular, we include a flat, 1\% uncorrelated systematic uncertainty in all bins when deriving expected limits and projections (unless otherwise specified). We also adopt a conservative statistical procedure by floating the normalization of the Drell-Yan distribution and fitting only to its shape, effectively dropping the impact of correlated systematics in our expected reach estimates. Moreover, we derive our limits using Drell-Yan distributions over logarithmically spaced mass bins, which should make our conclusions less dependent on resolution degradation at very high transverse momentum.

\begin{figure}[t]
\includegraphics[width=\textwidth]{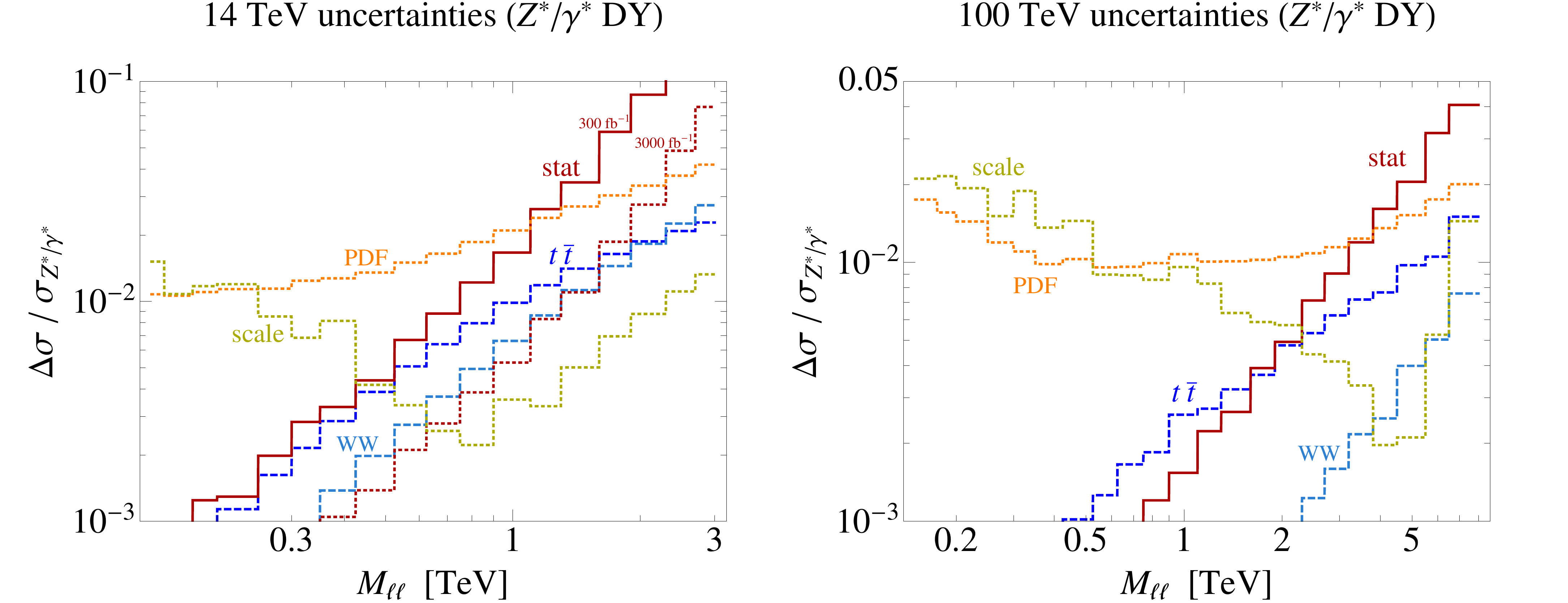} 
\caption{\small{Uncertainties in the dilepton invariant mass spectrum of neutral current DY at 14 TeV (left) and 100 TeV (right).  The statistical (red) and theoretical uncertainties - PDF (orange) and scale (yellow) - are obtained at fixed order NNLO with NNPDF2.3. Uncertainties from $t \bar t$ (dark blue) and $WW$ (light blue) backgrounds are assumed to be dominated by statistics in the $e^\pm\mu^\mp$ control region. At 14~TeV, background uncertainties are shown for 300 $\text{fb}^{-1}$, while statistical uncertainties are shown for 300 $\text{fb}^{-1}$(solid red) and 3000 $\text{fb}^{-1}$(dotted red). }
\label{fig:ErrDy}}
\end{figure}

\begin{figure}[t]
\includegraphics[width=\textwidth]{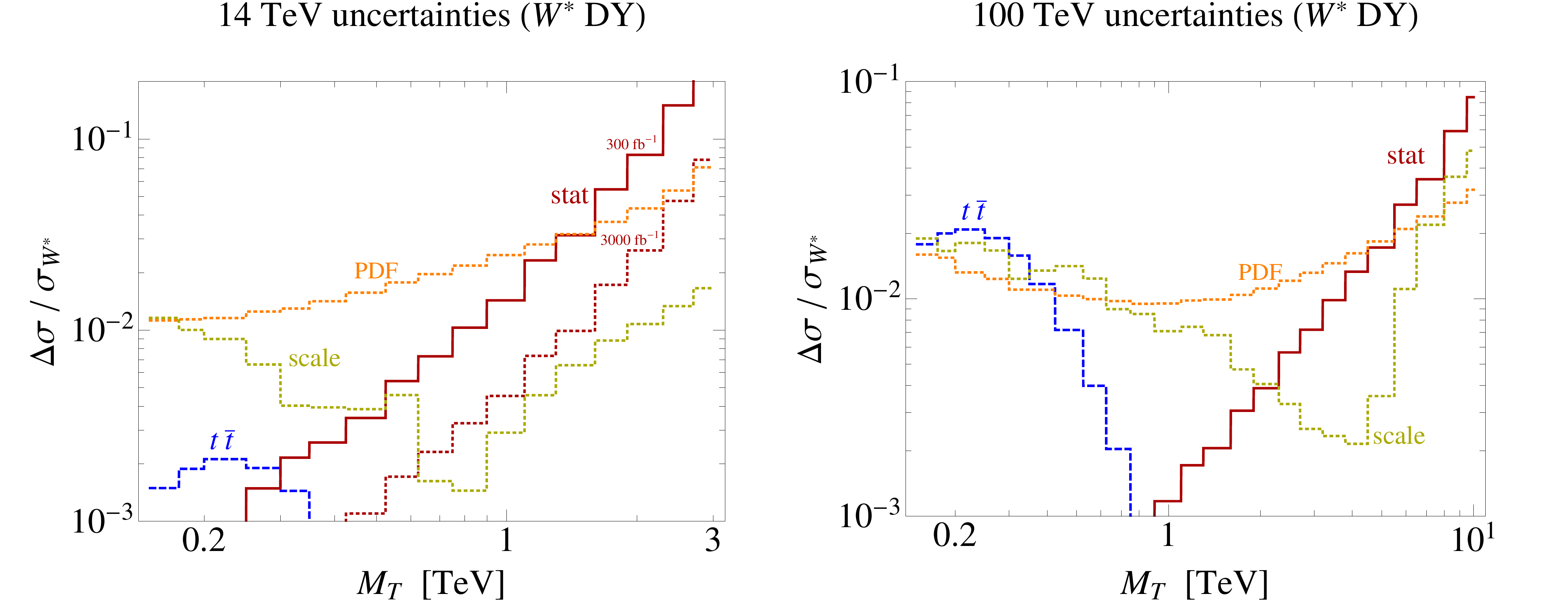} 
\caption{\small{Uncertainties in the transverse mass spectrum of charged current DY at 14 TeV (left) and 100 TeV (right).  The statistical, PDF and scale uncertainties are as in  \fig{ErrDy}. The $WW$ background is negligible and not shown. Uncertainties from $t \bar t$ subtraction are shown in blue, assuming a $b$-veto is applied and the $t \bar t$ cross section is known to 5\% precision.}
\label{fig:ErrW}}
\end{figure}

In addition to our assumption of systematic uncertanties being under control, we need to check that the dominant irreducible backgrounds at 14~and~100~TeV can be subtracted without introducing larger than percent-level uncertainties. We have done this with MC simulations of parton-level $t\bar{t}$ and $WW$ events that contribute to Drell-Yan final states. Diboson and $t\bar{t}$ contribute to $\ell^+\ell^-$ final states when the tops/$W$'s decay to same-flavor leptons. Their contribution, however, can be measured in control regions with opposite sign and opposite flavor (OSOF) lepton pairs (i.e., $e^+\mu^-$ and $e^-\mu^+$), and subtracted from the signal region. We assume that statistical uncertainties in OSOF measurements dominate this background subtraction, and display their contribution to $Z^*/\gamma^*$ uncertainties at 14~and~100~TeV in \fig{ErrDy}. We also show the Drell-Yan statistical and theoretical uncertainties. Both at 14~TeV and 100~TeV these irreducible backgrounds are negligible in the region of interest, and only become $\gtrsim \mathcal{O}$(1\%) where the statistical uncertainties are dominant and degrade sensitivity.

\begin{figure}[t]
\begin{center}
\includegraphics[width=0.6\textwidth]{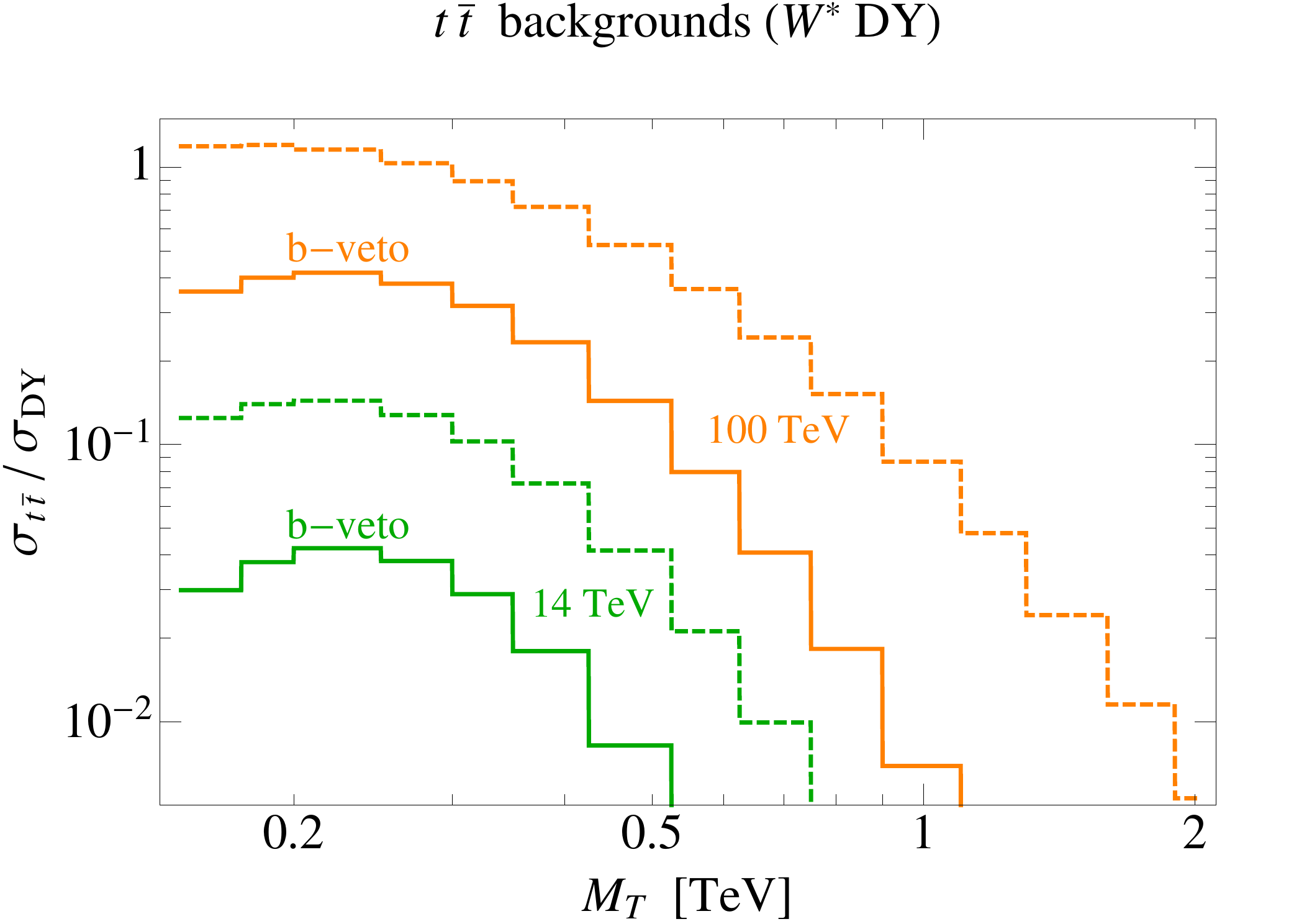} \quad
\caption{\small {The background to the transverse mass spectrum from $t \bar t$, at 14 and 100 TeV\@.  The background top cross section is normalized to the charged current DY cross section, and is shown before and after including a veto on $b$-jets.  We find that the background from $W^+W^-$, which is not shown, is significantly smaller than the background from tops.}
\label{fig:DYBG}
}
\end{center}
\end{figure}

Semi-leptonic $t\bar{t}$ and $WW$ events should not contribute substantially to high transverse mass $W^*$ Drell-Yan, since the lepton and missing energy (neutrino) originate from an on-shell $W$. However, dileptonic $t\bar{t}$ and $WW$ events do contribute to high $m_T$ bins when one of the leptons is missed (for instance when it falls out of acceptance). We have found that contamination from $WW$ background events is small both at 14~and~100~TeV (1 -- 2\% for $m_T \lesssim 500~\GeV$) and can be safely subtracted. The contribution from $t\bar{t}$, however, is comparable to the signal, as displayed by the dashed lines in \fig{DYBG}. At 14~TeV, the $t\bar{t}$ contamination to $W^*$ Drell-Yan is as large as $\mathcal{O}$(10\%) for $m_T \lesssim 400~\GeV$, and at 100~TeV it is $\mathcal{O}$(1) for $m_T \lesssim 500~\GeV$. Tagging b-jets should be a safe way to suppress $t\bar{t}$ backgrounds while keeping the Drell-Yan measurement inclusive and the theoretical uncertainties under control. Assuming an efficiency of 80\% for tagging b-jets with $p_T > 50~\GeV$ and $|\eta|<2.5$, the $t\bar{t}$ contamination can be reduced by a factor of $\sim3$, as displayed by the solid lines in \fig{DYBG}, which safely suppresses it to sub-percent levels at 14~TeV\@. At 100~TeV, the $t\bar{t}$ contribution still needs to be subtracted, and the residual uncertainty to the $W^*$ Drell-Yan measurement can be reduced to $<\mathcal{O}$(1\%) if the $t\bar{t}$ differential cross section is known to within 5\%. This is the present level of theoretical uncertainty for the total inclusive QCD NNLO $t\bar{t}$ cross section at 100~TeV, as obtained with {\tt Top++}\cite{Czakon:2011xx, Czakon:2013goa}, using NNPDF2.3: $\sigma_{t\bar{t}}^{\text{NNLO}}=35\hspace{0.1cm}716.9~\text{pb}^{~+2.87\% ~+0.88\% }_{~-4.74\%~-0.86\%}$, where scale and PDF uncertainties are quoted respectively. Differential $t\bar{t}$ cross sections at NNLO are being presently computed and should be known by the time such $\sqrt{s}=100~\TeV$ measurements are made. Even though we consider 5\% theoretical uncertainties a realistic expectation, we can more conservatively assume that the differential $t\bar{t}$ cross section will be known to within 10\%, in which case the $t\bar{t}$ subtraction will induce uncertainties $\mathcal{O}$(4 -- 5\%) for low mass bins $m_T\lesssim 500~\GeV$. We have checked that including this additional contribution to correlated systematics in the $m_T\lesssim 500~\GeV$ region causes a negligible deterioration in our expected reach and does not change our conclusions. \fig{ErrW} shows the contribution from $t\bar{t}$ subtraction to the uncertainties in $W^*$ Drell-Yan at both 14~TeV and 100~TeV, along with $W^*$ Drell-Yan theoretical and statistical uncertainties.

\subsection{Unknown Unknowns}
\label{subsec:unknowns}

In predicting and measuring cross sections, an important factor to control for is the presence of ``unknown unknowns'', aspects of the calculation/measurement that can be numerically significant but not properly accounted for\footnote{For the case at hand, one may consider the NNLO QCD corrections a ``known known'' and their scale uncertainty (an estimate of the higher order contributions) a ``known unknown''.  We leave the study of possible ``unknown knowns" for future work.}.  While Drell-Yan production is a well-understood process, the high mass regime far above the weak scale is one that colliders have only recently been able to probe, so that some consideration is required.

The high mass regime significantly expands the available phase space for radiation, leading to a broad spectrum of emissions in the inclusive cross section.  If the radiation is restricted, for instance through a veto on jets, this can lead to large logarithmic corrections that must be resummed.  Jet vetoes in QCD are well understood, and resummation may be used to control their perturbative uncertainties (see, e.g., Refs.~\cite{Banfi:2012jm,Becher:2013xia,Stewart:2013faa}).  A similar effect holds for EW contributions.  Virtual EW corrections are known to have a large effect at large scales $s \gg \MZ^2$, and their contribution will be largely mitigated for a measurement inclusive over EW radiation.  Vetoes on soft leptons, therefore, can introduce large EW corrections that may be difficult to control in theory predictions.  Because the initial state does not form an electroweak singlet, even the inclusive cross section contains EW Sudakov logarithms of the form $\ln s / \MZ^2$, and in the cross sections we study these logarithms will manifest as $\ln \Mll^2 / \MZ^2$ and $\ln \MT^2 / \MZ^2$.  Although we can calculate the NLO EW logarithms from virtual weak bosons and real and virtual photons (see \fig{ZprodKfac}), and resum the virtual corrections, understanding the interplay between real and virtual corrections and the all-orders properties of these effects is important to get the most precise predictions in the high mass regime of the cross section.  In \subsec{theory} we proposed an observable, $\MT^{\rm EW}$, that treats charged leptons and neutrinos equivalently and combines the contributions from charged and neutral current Drell-Yan channels.  This should be much less sensitive to EW Sudakov logarithms than $\Mll$ or $\MT$, while additionally allowing for a single measurement that combines the power of both Drell-Yan channels; further study is warranted.

A subset of the EW corrections are the contributions from a photon PDF, which may be numerically important for the neutral current Drell-Yan (and less important for the charged current case).  While the EW Sudakov logarithms may be calculated and resummed, the photon PDF ultimately must be measured, and measured at the LHC\@.  This creates a degeneracy in the measurement: if the Drell-Yan $\Mll$ spectrum is used to measure both the photon PDF and running EW couplings, then one would only simultaneously constrain both.  However, several features may help dissociate the two.  The functional dependence on the photon PDF and running EW couplings is different, and measurements across collider center-of-mass energies may be combined to further distinguish the two contributions.  Additionally, the charged current channel is not very sensitive to the photon PDF (for example, compare figures 23 and 24 of Ref.~\cite{Ball:2013hta}), meaning that the running of $\alpha_2$ and photon PDF may be simultaneously measured by considering both charged and neutral current channels, assuming no additional running of $\alpha_1$.  Finally, since the photon PDF increases the theoretical prediction for the Drell-Yan cross section, neglecting this contribution leads to a more conservative limit for the running EW couplings.  This is the approach that we adopt here.

\begin{figure}[t]
\begin{center}
\includegraphics[width=\textwidth]{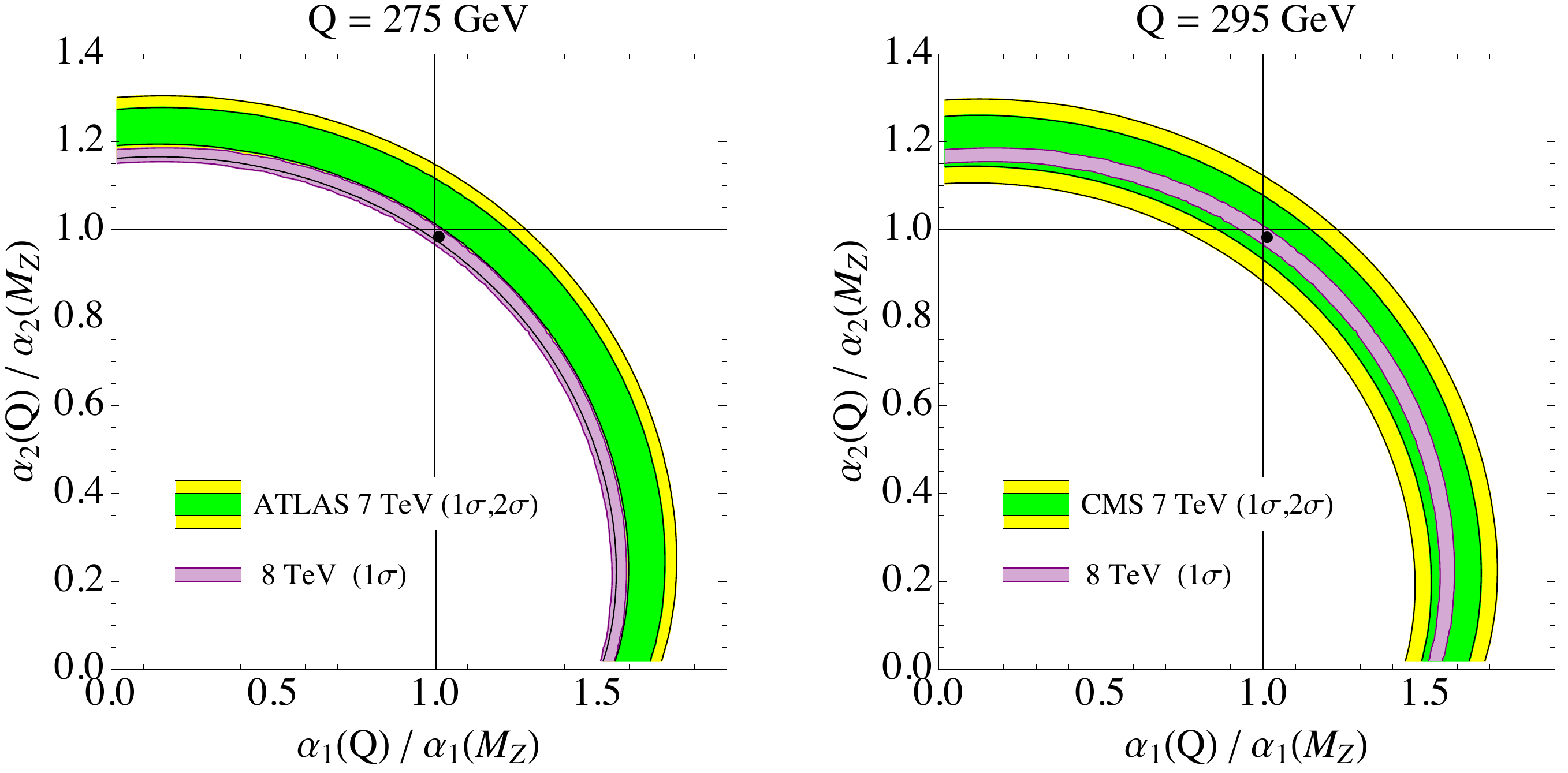}
\caption{\small {Fits to $\alpha_{1,2}(Q)$ using  7 TeV $Z^*/\gamma^*$ Drell-Yan data are shown for ATLAS at $Q = 275~\GeV$ (left) and for CMS at $Q = 295~\GeV$ (right). The $1\sigma$ ($2\sigma$) bands are shown in green (yellow) and the projected reach with 8 TeV data is shown in purple.  The SM prediction is given by a black dot.}}
\label{fig:RainbowCurrent}
\end{center}
\end{figure}

\section{Limits and Reach}
\label{sec:limits}

In this section we discuss current limits and future reach on the running of the electroweak couplings using Drell-Yan measurements, focusing on capabilities of hadron colliders operating at 8, 14, and 100 TeV, collecting 20, 300, and 3000 fb$^{-1}$ of data respectively. We refer the reader to \app{stats} for a detailed discussion of the statistical procedure we use for setting limits.

\subsection{Constraining $\alpha(Q)$}
\label{sec:alphaQ}

The measurement of the $Z^*/\gamma^*$ Drell-Yan cross section at an invariant mass bin centered around $Q=\Mll$ is a direct probe of the values of $\alpha_1$ and $\alpha_2$ at scale $Q$. As explained in Sec.~\ref{subsec:Zgamma}, we have inferred how the neutral current Drell-Yan cross section $\sigma_{Z^*/\gamma^*}(Q)$ scales with the electroweak gauge couplings (at leading order in those couplings). This allows us to place a limit on the deviation of $\alpha_{1,2}(Q)$ from their values at $\MZ$, $\alpha_{1,2}(\MZ)$. In order to factor out correlated systematic uncertainties, we can fit the ratio $\sigma(Q) / \sigma(Q_0)$, with $\sigma(Q_0)$ being the cross section in the first invariant mass bin above $\MZ$, where the logarithmic running of $\alpha_{1,2}$ can be neglected. This is most straightforwardly implemented in the profile likelihood method described in \app{stats} by including two bins only, at $Q_0$ and $Q$, and treating the relative normalization of their cross sections as a nuisance parameter.

\begin{figure}[b]
\begin{center}
\includegraphics[width=\textwidth]{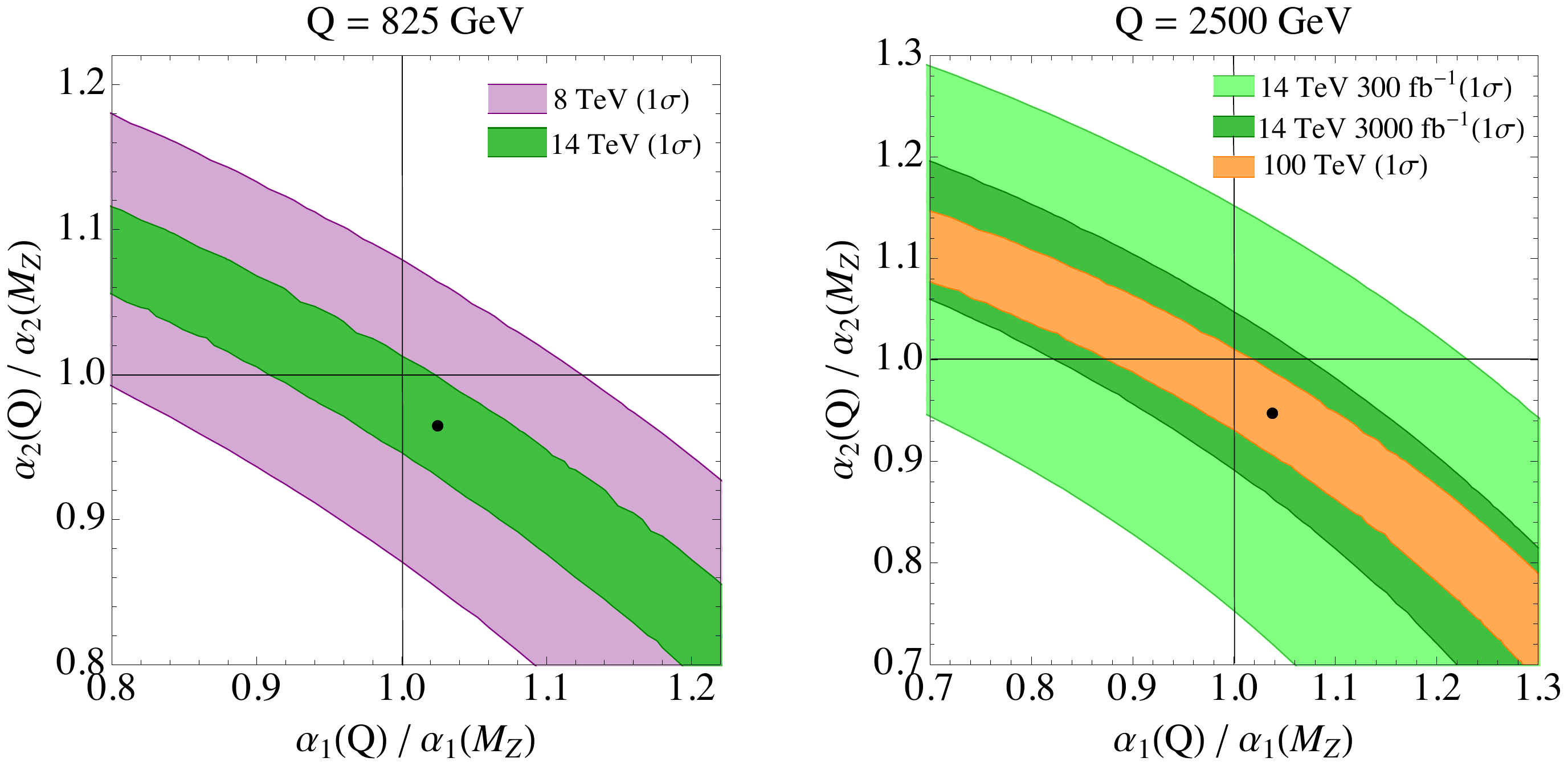}
\caption{\small {Projections of the reach to fit to $\alpha_{1,2}(Q)$ with future $Z^* / \gamma^*$ measurements, with $Q = 825$~GeV (2.5 TeV) on the left (right).  On the left, purple (green) show the $1\sigma$ reach from 8 (14) TeV running with a luminosity of 20 (300) fb$^{-1}$.  On the right, green (orange) show the $1\sigma$ reach at 14 (100) TeV\@.  Light (dark) green assumes a luminosity of 300 (3000) fb$^{-1}$ and the 100 TeV reach assumes a luminosity of 3000 fb$^{-1}$.  The SM prediction is given by a black dot.}}
\label{fig:RainbowFuture}
\end{center}
\end{figure}

In \fig{RainbowCurrent}, we use existing 7~TeV $Z^*/\gamma^*$ Drell-Yan measurements at the LHC~\cite{Aad:2013iua,Chatrchyan:2013tia} to place constraints on $\alpha_1$ and $\alpha_2$ at $Q= 275~\GeV$ and at $Q= 295~\GeV$ using ATLAS and CMS data, respectively. (The slight difference in $Q$ is due to different invariant mass binning between ATLAS and CMS.) CMS has made public their 8 TeV measurements of the neutral Drell-Yan differential cross section~\cite{CMS:2014hga}, but not the associated experimental uncertainties and their correlations. Without such information we cannot estimate existing 8 TeV limits on $\alpha_{1,2}$, so in \fig{RainbowCurrent} we display the expected 8 TeV improvement on the measurement of $\alpha_{1,2}$ at $Q= 275~\GeV$ and $295~\GeV$. \fig{RainbowFuture} displays the expected sensitivity on $\alpha_{1,2}(Q)$ at an intermediate scale ($Q=825~\GeV$) and a high scale ($Q=2.5~\TeV$). 

\begin{figure}[t]
\begin{center}
\includegraphics[height=7.0cm]{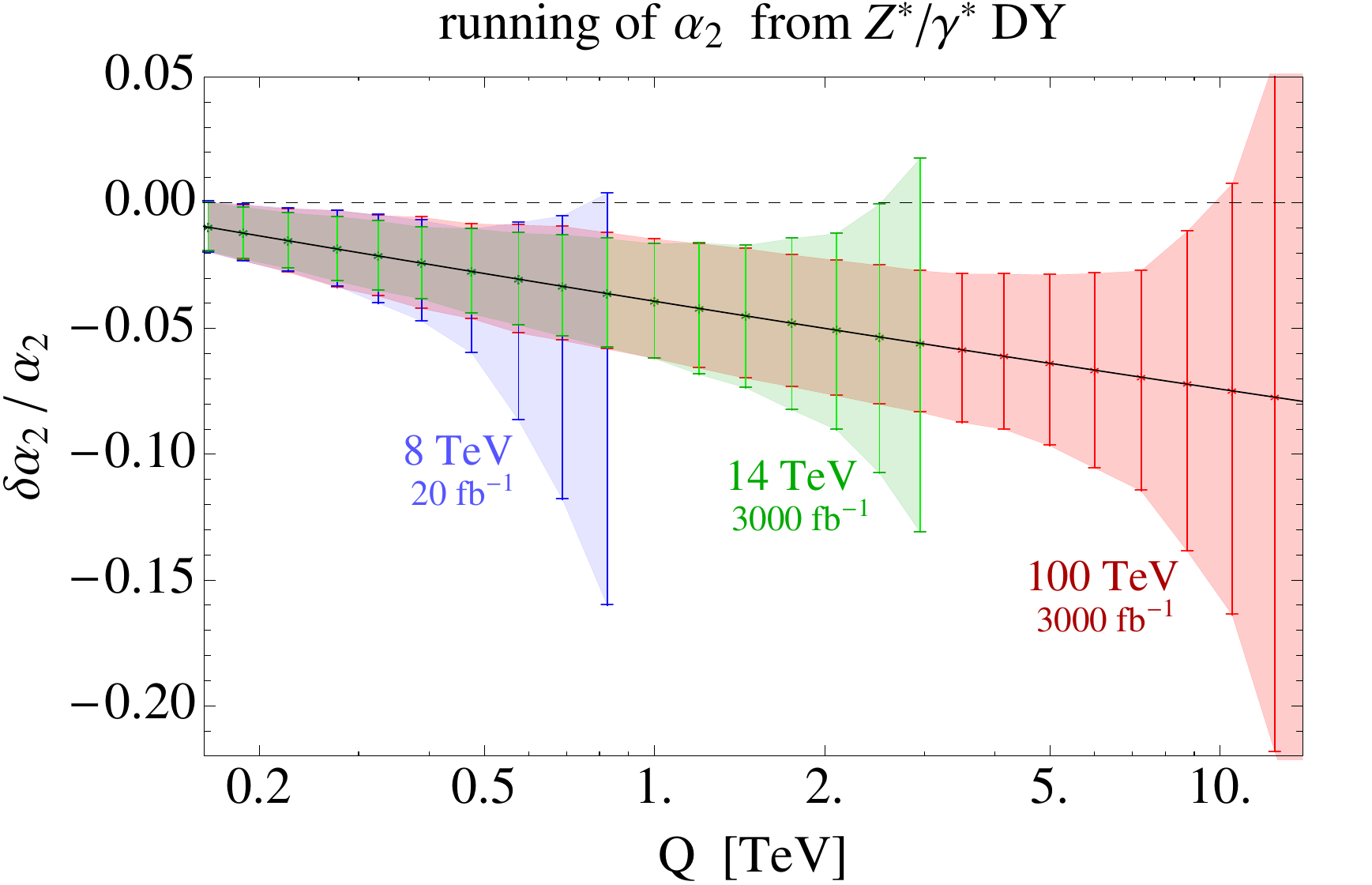}
\caption{\small {The projected reach to measure $\alpha_2(Q)$ with future $Z^* / \gamma^*$ measurements, assuming $\alpha_1$ runs according to the SM prediction.  Blue, green, red show the reach with 8, 14, and 100 TeV, assuming luminosities of  20, 3000, and 3000 fb$^{-1}$.  The error bars are $1 \sigma$.  The SM prediction (solid,black) is well-separated from the case that $\alpha_2$ does not run above $\MZ$ (dashed, black).}}
\label{fig:DeltaAlpha2vsQ}
\end{center}
\end{figure}

It is clear from \figs{RainbowCurrent}{RainbowFuture} that $\sigma_{Z^*/\gamma^*}(Q)$ is only sensitive to a combination of $\alpha_1$ and $\alpha_2$ at the scale $Q$. Measurements of $W^*$ Drell-Yan will break this degeneracy since $\sigma_{W^*}(Q)$ depends to leading order on only $\alpha_2$. This cannot be done, however, without further assumptions on the full scale dependence of $\alpha_2$, since the transverse mass bins receive contributions from all $W^*$ processes with $m_{W^*}\geq m_T$, so deviations of $\alpha_2$ at higher scales $Q^\prime> Q$ affect the cross section at $m_T \sim Q$.

Besides placing simultaneous limits on $\alpha_1$ and $\alpha_2$, one can assume that $\alpha_1$ runs as predicted in the Standard Model, and use the neutral current Drell-Yan measurements to constrain the running of $\alpha_2$. \fig{DeltaAlpha2vsQ} shows the expected bin-by-bin sensitivity on $\alpha_{2}$ for two decades in energy, $100~\GeV\lesssim Q\lesssim 10~\TeV$. For invariant mass bins below $\sim\!400~\GeV$, measurements at $\sqrt s = $ 14 and 100~TeV do not lead to a substantial improvement in sensitivity over 8~TeV because in that range PDF and uncorrelated systematic uncertainties are the limiting sensitivity factors. Above $Q\sim 400~\GeV$, statistical uncertainties become dominant in 8 TeV Drell-Yan, so the higher statistics at 14~TeV leads to an improvement in reach. Similarly, measurements at $\sqrt s = $ 100~TeV only supersede those at 14~TeV for $Q\gtrsim 1~\TeV$. Our projections indicate that there is sensitivity to discriminate the SM running of $\alpha_{2}$ from no running, and it should be possible to determine the sign of the SU$(2)_L$ beta function with high significance (as we will discuss shortly).

At the time of the writing of this paper, no Standard Model analyses of high-transverse mass $W^*$ Drell-Yan have been made public. ATLAS and CMS have looked at the high transverse mass distribution of final states with a lepton and missing transverse energy (MET) in Beyond the Standard Model (BSM) searches for $W'$ particles~\cite{Aad:2012dm, Chatrchyan:2013lga, ATLAS:2014wra, Khachatryan:2014tva}, but just as for the 8 TeV neutral Drell-Yan results, not enough information is publicly available for us to estimate existing limits on the running of the electroweak couplings with $\ell^\pm$+MET data (in particular, the correlation matrix for errors on the transverse mass spectrum across different bins would be necessary).

\begin{figure}[t]
\begin{center}
\includegraphics[width=\textwidth]{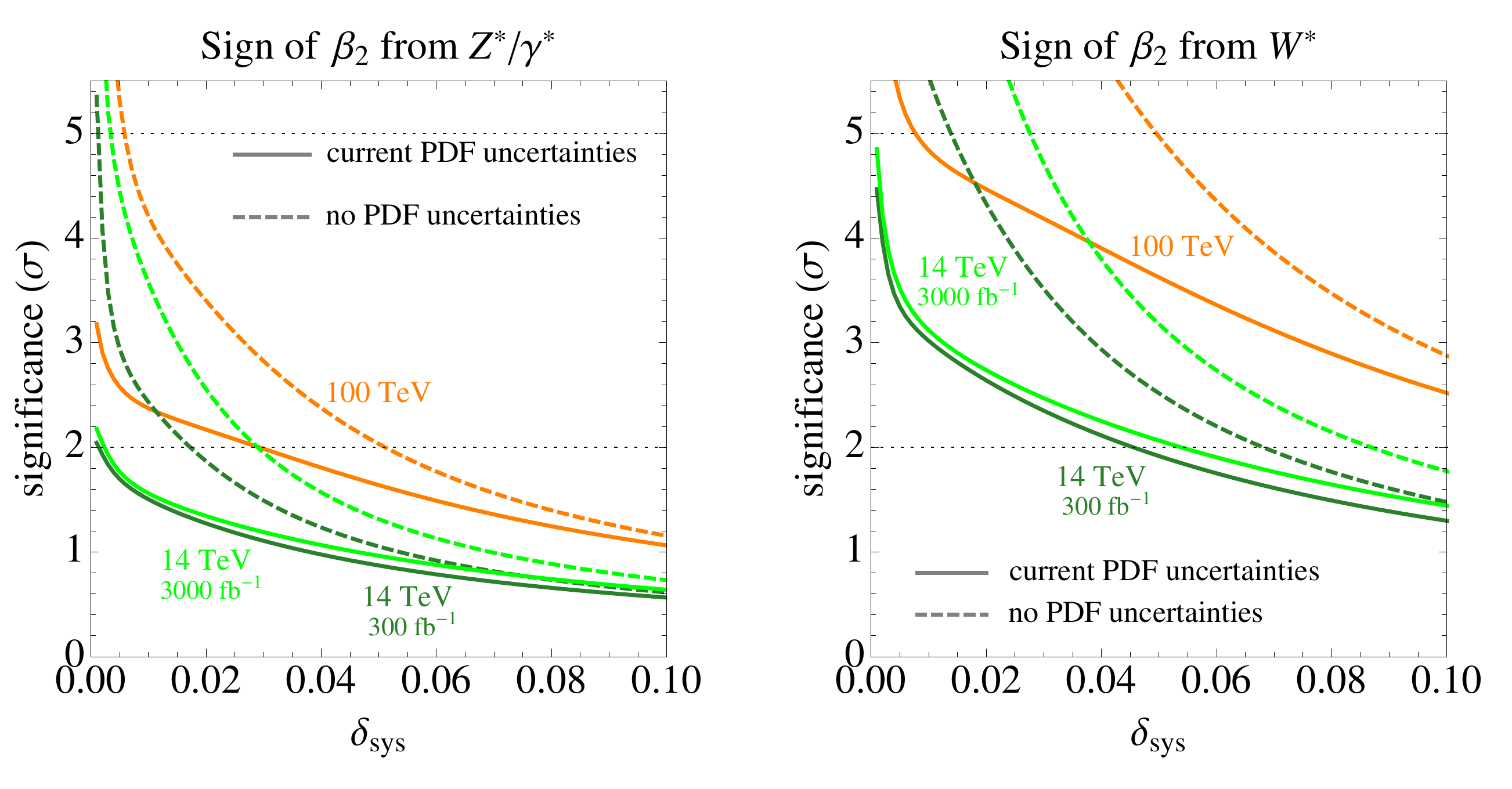}
\caption{\small {Projected exclusion of the no running hypothesis of $\alpha_2$ in $Z^* / \gamma^*$ (left) and $W^*$ (right) Drell-Yan, as a function of the size of uncorrelated systematic uncertainties. Expected limits are shown assuming present level PDF uncertainties (solid curves) as well as negligible ones (dashed curves), for 14 TeV (green) and 100 TeV (orange).}}
\label{fig:signBeta2}
\end{center}
\end{figure}

\subsection{Constraining New Physics Scenarios}

In the previous section we discussed how to constrain $\alpha_{1,2}$ at a given scale $Q$ by measuring the neutral current Drell-Yan cross section at a bin centered around $Q$. When constraining BSM scenarios that modify the running of the electroweak couplings, however, the strongest exclusions come from combining all bins, both in $Z^*/\gamma^*$ and $W^*$ measurements. In this section the projections we will derive will be of that sort: given a prediction for the running of $\alpha_{1,2}$, we will estimate its exclusion at 8,~14~and~100~TeV colliders from the entire high mass Drell-Yan distribution ({\it i.e.}, above the $Z$ peak), assuming that observations are consistent with the running predicted in the Standard Model.  Since we are using the leading order EW coupling dependence to derive the limits on BSM scenarios, the reach curves we show are sensitive to the scale choice for the couplings.  Our limits show the potential reach, but we leave a full study of NLO EW effects with new physics to future work; see \app{EWscale}.

\begin{figure}[t]
\begin{center}
\includegraphics[width=\textwidth]{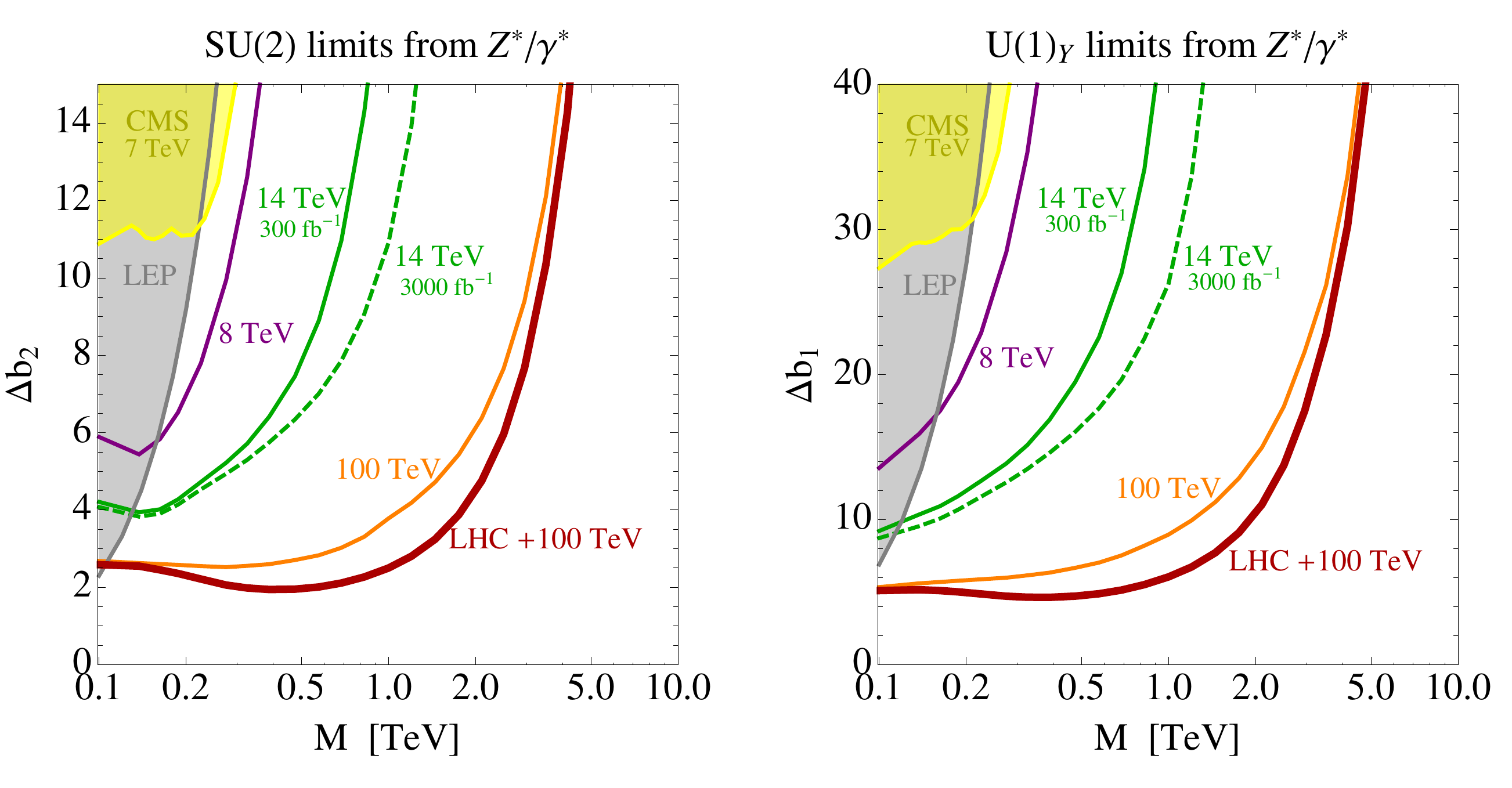}
\caption{\small {$Z^*/\gamma^*$ 2$\sigma$ reach to BSM contributions to the EW beta function coefficients, $\Delta b_2$~(left) and $\Delta b_1$~(right), as a function of the mass of the new states, $M$. Current limits from LEP and 7~TeV~DY at CMS are indicated by shaded regions. Projected reach is shown for 8 TeV (purple), 14 TeV (green), 100 TeV (orange), and LHC+100 TeV (red). }}
\label{fig:BvsMneutral}
\end{center}
\end{figure}

We indicated in Sec.~\ref{sec:alphaQ} that the sign of the SU$(2)_L$ beta function could be well measured in Drell-Yan distributions. In \fig{signBeta2} we show the significance with which future measurements can exclude the hypothesis of no running of the SU$(2)_L$ gauge coupling as a function of the size of the uncorrelated systematic uncertainties. That is equivalent to measuring the sign of the SU$(2)_L$ beta function at the weak scale. We show the expected exclusion separately for neutral (left) and charged (right) Drell-Yan, at 14~TeV~ ($300~\text{fb}^{-1}$,~$3000~\text{fb}^{-1}$) and 100~TeV ($3000 ~\text{fb}^{-1}$). We also display the asymptotic expected reach in the case of improved measurements of proton PDFs, with its uncertainties reduced to negligible levels. The strongest exclusions come from $W^*$ Drell-Yan, and can have a significance of $3\sigma$ ($5\sigma$) at 14~TeV (100~TeV), if the uncorrelated systematics are $\lesssim 1\%$. If the PDF uncertainties were reduced to negligible levels at the region of interest for Drell-Yan, a $5\sigma$ exclusion would be possible at the LHC (100 TeV collider) for uncorrelated systematics as large as 3\% (5\%).

\begin{figure}[t]
\begin{center}
\includegraphics[width=\textwidth]{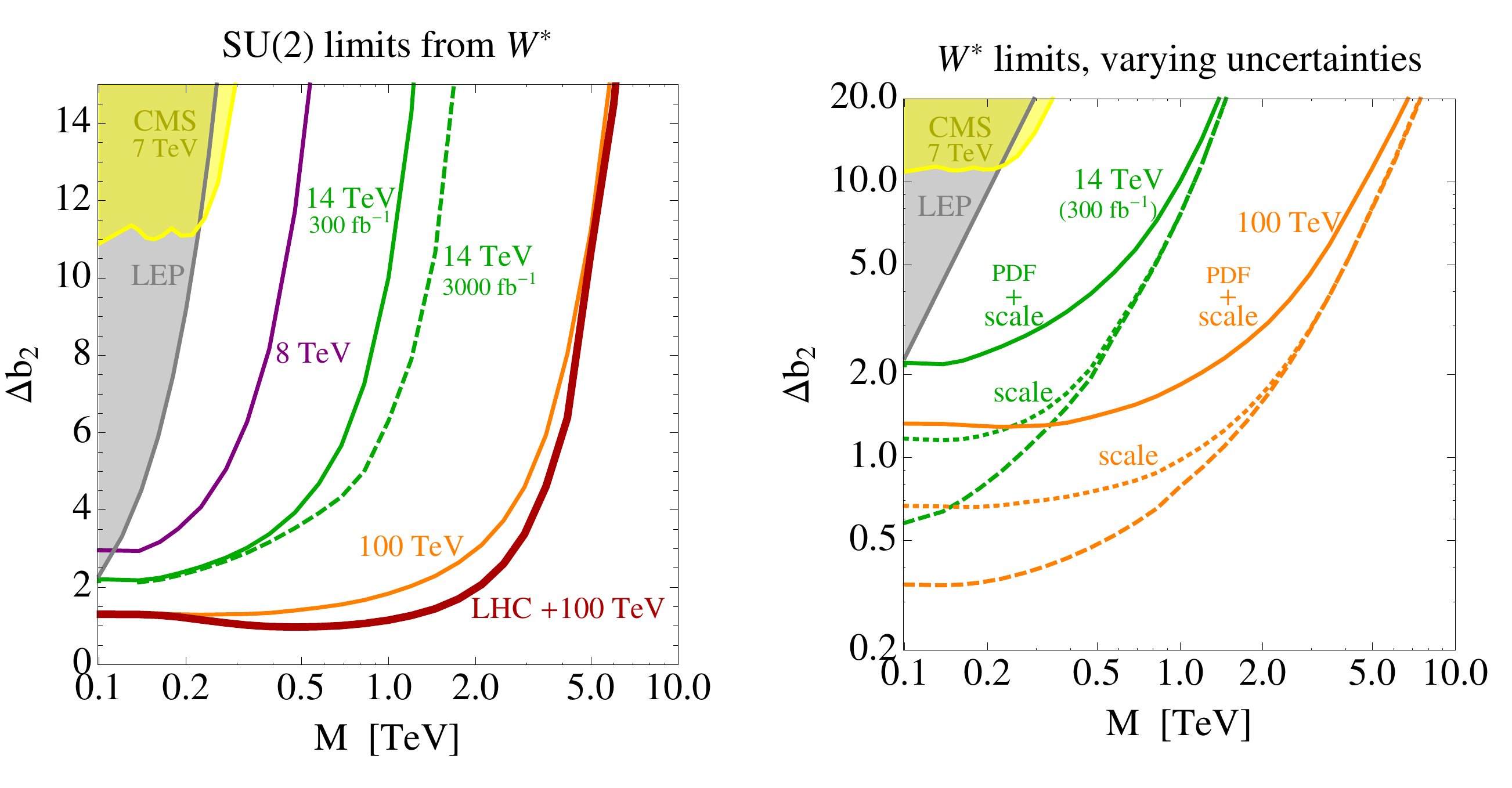}
\caption{\small {$W^*$ 2$\sigma$ reach to BSM contributions to the SU$(2)_L$ beta function coefficient, $\Delta b_2$, as a function of mass of the new states, $M$. Current limits from LEP and 7 TeV DY at CMS are indicated by shaded regions. Projected reach is shown for 8 TeV (purple), 14 TeV (green), 100 TeV (orange), and LHC+100 TeV (red). All (solid) curves on left (right) plot assume current PDF and scale uncertainties. Dotted curves in the right plot assume the presence of only scale uncertainties, and dashed curves show the result of removing theoretical uncertainties.}}
\label{fig:BvsMcharged}
\end{center}
\end{figure}

More generally, one can place bounds on the change of the EW beta function coefficients, $\Delta b_{1,2}$, as a function of scale, $M$. These can also be interpreted as mass/charge bounds on new electroweak states, since such states will contribute to the beta function coefficients $b_{1,2}$  above their mass threshold, with $\Delta b_{1,2}$ depending on their electroweak charges. \figs{BvsMneutral}{BvsMcharged} show the expected ($2\sigma$) reach in the $\Delta b_{1,2}$ - $M$ plane for various center of mass energies. In \fig{BvsMneutral}, neutral current Drell-Yan limits are shown, assuming that only one electroweak beta function is modified at a time (namely, $\beta_2$ on left and $\beta_1$ on right) while the other remains as predicted in the Standard Model. Projections at 8~TeV~($20~\text{fb}^{-1}$), 14~TeV~($300~\text{fb}^{-1}$,~$3000~\text{fb}^{-1}$) and 100~TeV~($3000~\text{fb}^{-1}$) are shown, as well as the combination of 8,~14~and~100~TeV measurements (which we refer to as the LHC+100~TeV combination). We also show EW precision constraints from LEP on SU$(2)_L$ and custodial isospin-preserving states, as explained in Sec.~\ref{subsec:IRrunning}, and existing limits from CMS 7~TeV Drell-Yan measurements. (Limits from ATLAS 7~TeV are weaker due to an upward fluctuation in their data relative to the SM prediction, and do not appear in the plot.) \fig{BvsMcharged}(left) displays the charged current Drell-Yan expected reach for the $\Delta b_{2}$ - $M$ plane, which is computed assuming that the $W^{*+}$ and $W^{*-}$ distributions are separately measured. \fig{BvsMcharged}(right) shows the sensitivity improvement at 14~TeV ($300~\text{fb}^{-1}$) and 100~TeV if theoretical uncertainties were improved relative to present levels. Namely, the solid line is computed assuming present PDF+scale uncertainties (and are the same as in left plot), while the dotted lines assume present scale uncertainties but negligible PDF uncertainties. Finally, the dashed lines assume negligible PDF+scale uncertainties. The actual BSM reach from these measurements should lie between the solid and dashed curves in \fig{BvsMcharged}(right), assuming that LHC data will improve PDF and scale uncertainties.

In \figs{BvsMneutral}{BvsMcharged} we see that combining different $\sqrt{s}$ data (namely, 8,~14~and~100~TeV) leads to a non-negligible improvement of reach in the mass range $300 ~\GeV \lesssim M \lesssim 4~\TeV$ over the reach from 100~TeV data alone. Given that the sensitivity at 8~TeV and 14~TeV degrade around $M\sim 500~\GeV$ and $M\sim 1~\TeV$, respectively, the improvement from the combination does not come from a reduction in the uncorrelated uncertainties, but from the fact that we take into account the correlations in the PDF and scale uncertainties across different center-of-mass energies. In effect, measuring the Drell-Yan rate at a given invariant mass bin at different $\sqrt{s}$ probes the same scale for the electroweak couplings but different x-regions of the proton PDFs, essentially disentangling running couplings effects from PDF uncertainities (including contamination from photon initiated processes). This provides a good motivation for running a future proton-proton machine at different center-of-mass energies.

\begin{figure}[t]
\begin{center}
\includegraphics[width=\textwidth]{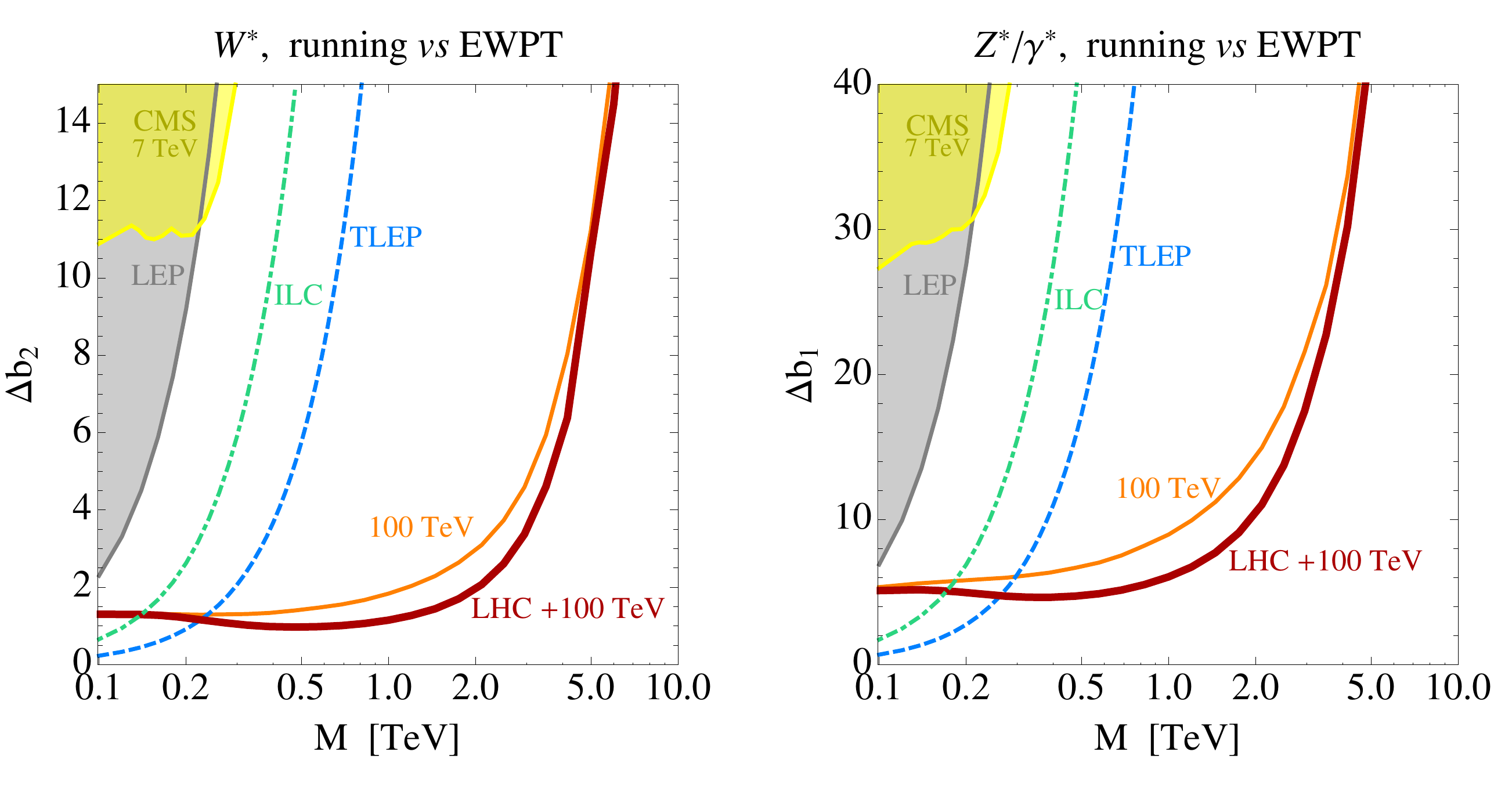}
\caption{\small {Contrast in reach between EWPT and DY measurements (left: $W^*$, right: $Z^*/\gamma^*$), to BSM contributions to the electroweak beta function coefficients, $\Delta b_2$~(left) and $\Delta b_1$~(right), as a function of mass, $M$. Current limits from LEP (shaded gray) and 7 TeV DY at CMS (shaded yellow) are shown as well. }}
\label{fig:BvsWY}
\end{center}
\end{figure}

Electroweak precision tests can strongly constrain new electroweak states, especially those that generate SM four-fermion operators at tree-level and contribute to the breaking of EW and/or custodial symmetry (thus generating the $S$ and/or $T$ parameters),  but their bounds are fairly mild on states that couple only through weak gauge charges and preserve these symmetries. For the latter case, Drell-Yan measurements at the 7 TeV LHC are already competitive with LEP bounds, as indicated by the shaded regions in Figs.~\ref{fig:BvsMneutral}-\ref{fig:BvsWY}, and will surpass LEP at 8~TeV\@. In \fig{BvsWY}, we contrast  Drell-Yan with EWPT projections. The Drell-Yan reach is shown for 100~TeV and the LHC+100~TeV combination, while the EWPT reach is shown for ILC and TLEP\@. In the very low mass region both measurements are competitive. In particular, EWPT have better sensitivity to modifications in the running of the hyperchage coupling below 300 GeV\@. In contrast, EWPT sensitivity deteriorates dramatically in the moderate to large mass region $M\gtrsim 500~\GeV$, which can only be probed by Drell-Yan measurements at 100 TeV  .

\begin{figure}[t]
\begin{center}
\includegraphics[width=\textwidth]{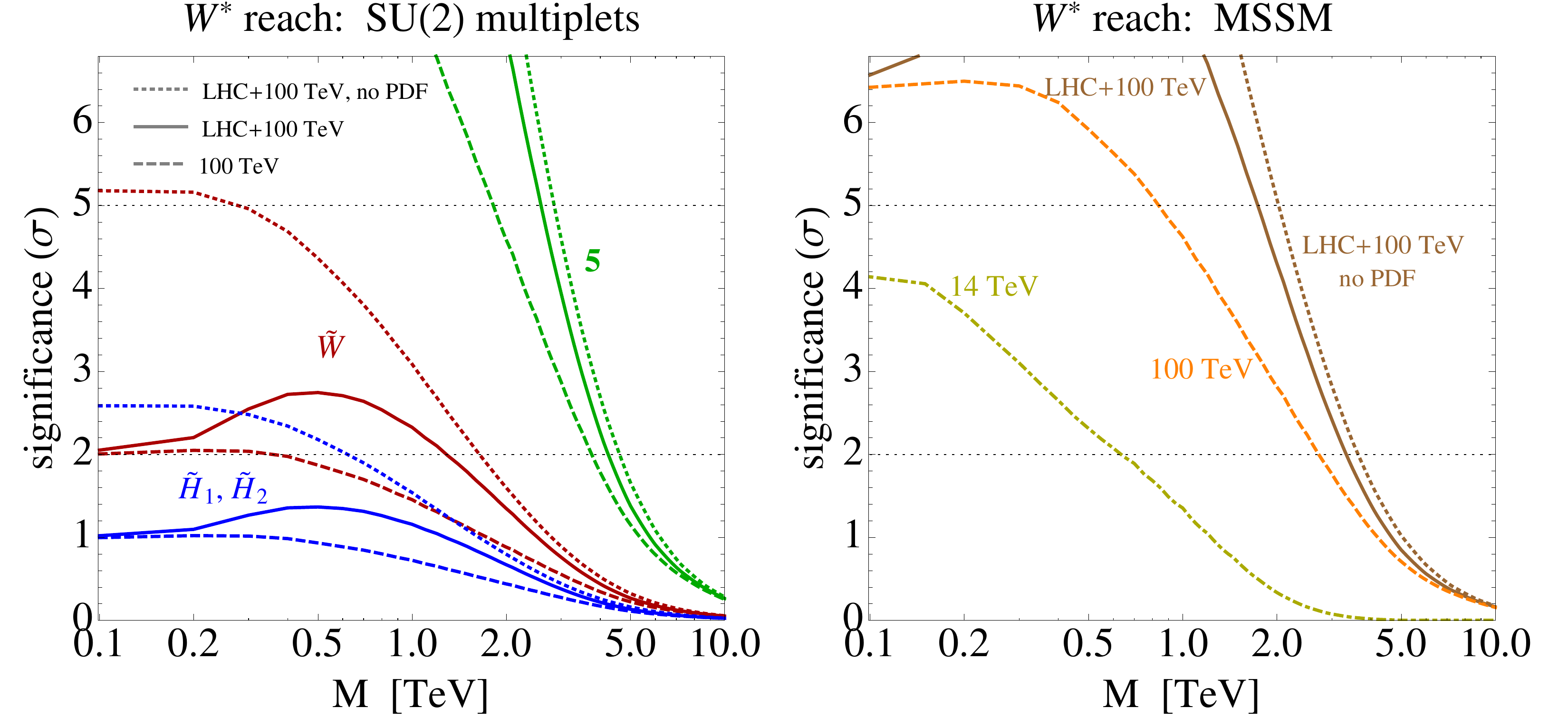}
\caption{\small {$W^*$ expected exclusion (in standard deviations) as a function of mass for various SU$(2)_L$ representations. {\it Right}: Sensitivity is shown at 100~TeV and LHC+100~TeV for a pair of Higgsinos (blue), a wino (red) and a 5-plet of SU$(2)_L$ (green). {\it Left}: Sensitivity to the MSSM running of SU$(2)_L$ is shown at 14,~100~and~LHC+100~TeV assuming that the contribution from all SU$(2)_L$-states enter at the same mass, $M$.}}
\label{fig:SU2multiplets}
\end{center}
\end{figure}

Finally, in \fig{SU2multiplets} we show the $W^*$ Drell-Yan reach for specific SU$(2)_L$ representations. The left plot displays expected limits for a fermionic triplet (wino), a pair of fermionic doublets (Higgsinos), and a fermionic quintuplet of SU$(2)_L$. While the former two appear in the context of supersymmetry, the later is motivated by minimal dark matter (MDM) scenarios~\cite{Cirelli:2005uq, Cirelli:2007xd}. On the right plot, the sensitivity is shown for the MSSM running of $\alpha_2$, assuming for the sake of illustration that the contribution to $\Delta b_2$ from all SU$(2)_L$ states enter at the same mass threshold, $M$. The projections are shown as expected exclusion (in number of standard deviations) versus mass. The left plot shows the reach for 100~TeV and the LHC+100~TeV combination, as well as the improved LHC+100~TeV reach when PDF uncertainties can be neglected. Note, in particular, that a wino with mass $\lesssim 1.3~\TeV$ could be excluded at 2$\sigma$ with combined LHC+100~TeV Drell-Yan data. That is competitive with reach projections from monojet searches at 100~TeV~\cite{Low:2014cba}, but less sensitive than (the more model-dependent) searches for disappearing tracks. While a MDM quintuplet as heavy as ($m_{\boldsymbol{5}}\sim 5~\TeV$) may be excluded with combined LHC+100~TeV Drell-Yan data, the reach for Higgsinos is modest, $m_{\tilde{H}} \lesssim 600~\GeV$, and requires a reduction of PDF uncertainties to negligible levels. The right plot displays the expected reach to the MSSM running for 14~TeV, 100~TeV, and the LHC+100~TeV combination assuming present as well as negligible PDF uncertainties. In particular, with this simplified (but conservative) spectrum hypothesis, the LHC+100~TeV combination could exclude the MSSM, at $2\sigma$, for $M_\text{MSSM}\lesssim 3.3~\TeV$, as well as discover it, at $5 \sigma$, for $M_\text{MSSM}\lesssim 1.8~\TeV$, where $M_\text{MSSM}$ can be conservatively interpreted as the heaviest SU$(2)_L$-charged MSSM state. Note that even though squarks in that mass range would be well within reach of the LHC and a future 100 TeV collider, exclusions from direct searches would be model dependent, and could {\it a priori} be evaded in RPV or stealth~\cite{Fan:2011yu,Fan:2012jf} scenarios, while still being probed by the running of electroweak couplings.

In the reach curves shown in this section, we have used the leading order EW coupling dependence of the Drell-Yan cross sections to derive the limit on BSM scenarios using measurements of the running EW couplings.  A higher order calculation of the Drell-Yan cross sections, including EW effects from the BSM matter content, is needed to provide the most robust theoretical predictions from which a measurement can be made.  While such predictions are beyond the scope of this work, our choice of the dilepton invariant mass for the central scale is well-motivated but also implies a nontrivial uncertainty on the reach curves as this scale is varied.  In \app{EWscale}, we show how the reach curves vary with the coupling choice for a couple of example models.  Precision EW calculations of the Drell-Yan cross sections will have small EW scale uncertainties and correspondingly significantly narrow the scale dependence of the reach curves.

\section{Conclusions}
\label{sec:conclusions}

The EW gauge couplings, $\alpha_{1,2}$, have been measured precisely at $Q = \MZ$, but their energy dependence above $\MZ$ is unconstrained.  In this paper, we have argued that high energy proton colliders can be used to measure running $\alpha_{1,2}$ above the weak scale. The shape of the lepton invariant mass spectrum in neutral current Drell-Yan, $\Zproc$, constrains $\alpha_{1,2}(Q)$ and the shape of the transverse mass spectrum in charged current Drell-Yan, $\Wproc$, constrains $\alpha_2(Q)$.  Measurements of running EW couplings are interesting Standard Model measurements, and can also be interpreted as model-independent searches for new states with EW quantum numbers.

Our main results include the following:
\begin{itemize}
\item  We have analyzed the statistical and systematic uncertainties that limit the precision with which running EW couplings can be measured.  The Drell-Yan cross section, which sets the statistical uncertainties, is shown in \figs{NevDY}{NevWtot}.  The systematic uncertainties coming from PDF, scale variation, and top and diboson background subtraction are summarized in \figs{ErrDy}{ErrW}.  We find that a precision better than $\sim 1-2 \%$ is achievable across a wide range of energies.
\item We use 7 TeV LHC data to perform the first fits to running EW couplings, shown in \fig{RainbowCurrent}.  Projected improvements on these fits for higher collision energies are shown in \fig{RainbowFuture}.
\item We project the reach to constrain new states with EW quantum numbers, at $\sqrt s = 8,14$, and 100 TeV, as shown in \figs{BvsMneutral}{BvsMcharged}.
We find that the 8 TeV dataset can now be used to set the strongest current model-independent limits on new particles that couple through electroweak quantum numbers,  surpassing electroweak precision tests.  
\item  At a future 100 TeV collider, we find $2 \sigma$ sensitivity to winos up to $\sim$ 1.3 TeV, a {\bf 5}-plet of SU(2) up to $\sim5$ TeV, and the $SU(2)$ doublets of the MSSM up to $\sim 3.3$ TeV, as shown in \fig{SU2multiplets}.  This reach may vary as the EW contributions from new physics is more precisely included in the limits, see \app{EWscale}.
\item Measurements of running $\alpha_2$ can determine the sign of $\beta_2$, testing the SM prediction of asymptotic freedom for $\alpha_2$.  The 14 TeV LHC (future 100 TeV collider) can measure the sign of $\alpha_2$ at $2-3 \sigma$ ($4-5\sigma$), as shown in \fig{signBeta2}.
\end{itemize}

Our study builds on previous work, which focused on SUSY corrections to Drell-Yan.  Ref.~\cite{Rainwater:2007qa} suggests using Drell-Yan to measure EW running and studies the sensitivity of the LHC to differentiate SM and MSSM running.  Refs.~\cite{Dittmaier:2009cr,Brensing:2007qm} study the full NLO EW corrections induced by the MSSM.  We improve upon the above studies by advocating a model-independent approach,
performing detailed analysis of the relevant uncertainties, and performing the first studies of Drell-Yan at the $\sqrt s = 100$~TeV energy scale.


We would like to highlight future steps that can help to achieve accurate measurements of  running EW couplings:
\begin{itemize}
\item The LHC should measure the high transverse mass spectrum of $W^*$, including a careful study of the correlation matrix among bins of different $m_T$ (analogous to the existing careful measurements of the dilepton invariant mass spectrum~\cite{Aad:2013iua,Chatrchyan:2013tia,CMS:2014hga}).
\item New states with EW quantum numbers should be included in NLO EW corrections to the Drell-Yan cross section.  The running gauge coupling captures the leading logarithmic effect, but the new states should be included in the EW loops in order to include finite corrections.  The simplified parameter space $(M, \Delta b_1, \Delta b_2)$ can be used.  This may be used to make more robust predictions for the reach of running EW coupling measurements for BSM models (see \app{EWscale}).
\item The EW Sudakov logarithms in the Drell-Yan $\Mll$ and $\MT$ cross sections should be further studied, and resummed, in the large $\Mll$ or $\MT$ regimes.  Such a study must account for both virtual corrections and real radiation, in order to control the theoretical uncertainties from EW corrections.  The level of incomplete cancellation between these corrections, due to the EW non-singlet initial states or the exclusivity of the event selection cuts, will affect the net EW correction in the high mass regime and feed into any uncertainty that should be assigned to the EW corrections.
\item The LHC should measure the photon PDF as precisely as possible, taking advantage of neutral current Drell-Yan observables such as the positively (or negatively) charged lepton transverse momentum to maximize sensitivity to the photon PDF~\cite{Dittmaier:2009cr,Boughezal:2013cwa}.
\end{itemize}

\vspace{0.2cm}
{\bf Acknowledgments} \\

We thank Kyle Cranmer, Mike Hance, Beate Heinemann, Ye Li, Michelangelo Mangano, Alex Mitov, Frank Petriello, Matt Reece, Filippo Sala, Daniel Stolarski, Tim Tait, and Liantao Wang for useful discussions.  JTR thanks the participants of the workshop ``BSM physics opportunities at 100 TeV," at CERN, for helpful comments.  JTR thanks the CERN theory group and the CFHEP at IHEP for hospitality while part of this work was completed.   JG thanks the Aspen Center for Physics for hospitality during the completion of this work, supported in part by National Science Foundation Grant No. PHYS-1066293.  The work of JRW was supported by the Director, Office of Science, Office of High Energy Physics of the U.S. Department of Energy under the Contract No. DE-AC02-05CH11231.  This work used resources of the National Energy Research Scientific Computing Center, which is supported by the Office of Science of the U.S. Department of Energy under Contract No. DE-AC02-05CH11231. DSMA is supported by the NSF under grants NSF-PHY-0969510 (the LHC Theory Initiative), PHY-0947827 and PHY-1316753.  JG is supported by the James Arthur Postdoctoral Fellowship at NYU.

\appendix

\section{Notation and Conventions for Beta Functions}
\label{app:beta}

At the leading log level, effects from new states above their mass threshold are concisely captured by gauge couplings' beta functions.  In general we can expand these functions in powers of the coupling:
\beq
\frac{dg}{d \ln \mu} = \frac{b^{(1)}}{16 \pi^2} g^3+ \frac{b^{(2)}}{(16 \pi^2)^2} g^5 + \mathcal O(g^7).
\eeq
We work to leading order in this expansion, as the couplings we are interested in are small and higher order effects contribute negligibly.  Thus we drop superscripts and define $b=b^{(1)}$ as the function under consideration.  
For a general $SU(N)$ gauge theory with matter content in representations $r$, this one-loop beta function is given by
\beq
b = -\frac{11}{3} N + \frac{2}{3} \sum_{f, \, r} T_r + \frac{1}{3} \sum_{s,\, r} T_r.
\eeq
The sums are over all representations of fermionic and scalar  fields, respectively.  
For abelian groups, a particle of charge $Q$ contributes as $T = Q^2$.  Contributions for non-abelian groups are dictated by the index defined via the generators $T_r^a$ of a given representation as
\beq
{\rm Tr} \, (T_r^a T_r^b) = T_r \delta^{ab}; \quad T_\square = 1/2.
\eeq

Finally, regarding renormalization scheme, we use mass-independent subtraction (e.g. $\overline{\rm MS}$) such that a particle contributes to   $b$ only at scales above its mass, and such that its contribution at $\mu > m$ is constant.  Below its mass, a particle must be manually integrated out in this scheme, with residual effects in IR appearing as irrelevant operators.  With a particle of mass $m$ contributing  $\Delta b$ to a given beta function, the corresponding gauge coupling measured at $\Lambda<m$ is thus determined at scales $\mu>\Lambda$:
\beq
\frac{1}{g^2(\mu)} = \frac{1}{g^2(\Lambda)} - \frac{b}{16 \pi^2} \ln \left( \frac{\mu^2}{\Lambda^2} \right) 
-\theta(\mu - m) \frac{\Delta b}{16 \pi^2}  \ln \left( \frac{\mu^2}{m^2} \right) ,
\eeq
with $b$ the contribution from any other light states in the spectrum.  In particular, the beta function is discontinuous across  mass thresholds.  The running of EW gauge couplings in the SM, for instance, is determined at scales
$\mu >m_t$ to be
\beq
b_1^{\rm (SM)} = \frac{41}{10}, \quad b_2^{\rm (SM)} = -\frac{19}{6}.
\eeq
Note that we appeal to a GUT normalization for the abelian factor where hypercharge is related to diagonal $SU(5)$ generators such that $g_Y = \sqrt{3/5} \times g_1$ and $b_1 = 3/5 \times b_Y$.

\section{Rescaling Function for Neutral Current Drell-Yan Cross Section}
\label{app:rescaling}
We parametrize the cross section $\frac{\df\sigma^{Z/\gamma}}{\df \Mll}$ in terms of the variables $\hat s_W$ and $a$ introduced in \eq{aalpha}; here we collect  expressions for the coefficients of the rescaling function that defines the general differential partonic cross section at leading order in the weak couplings.

As described in Sec.~\ref{subsec:Zgamma}, we write
\beq
\df  \sigma^{Z/\gamma} = a^2 f(\hat s_W^2, \Mll; Q) \times \df  \sigma_0^{Z/\gamma}, 
\eeq
with
\beq
f(\hat s_W^2, \Mll; Q) = c_0 +c_1 \hat s_W^2 + c_2 \hat s_W^4 + c_3 \hat s_W^6 + c_4 \hat s_W^8,  
\eeq
and proportionality constant
\beq
\df  \sigma_0^{Z/\gamma} = \frac{\pi \Mll^2}{2} \frac{1}{(\Mll^2-\MZ^2)^2+\MZ^2 \Gamma_Z^2} \times \df  \eta \, \df \Mll^2 \, \delta(\hat s - \Mll^2). 
\eeq
We define $c_i$ in terms of relevant kinematic parameters and functions of the lepton rapidity in the partonic CM frame:
\beq
h_1 = \frac{1+\tanh^2 \eta}{\cosh^2 \eta}\, ,  \qquad
h_2  = \frac{\tanh \eta}{\cosh^2 \eta}\, .
\eeq
We then have
\beq
c_0 &=& \left(h_1 + 2 h_2 \right) \times  \frac{1}{4} \, T_f^2 T_i^2  \\
c_1 &=& \left( h_1+2 h_2\right) \times  \frac{1}{2}  \left[ - Q_f T_f T_i^2 -  T_f^2 Q_i T_i  + Q_f T_f Q_i T_i \left(1- \frac{\MZ^2}{\Mll^2}\right) \right] \\
c_2 &=& h_1 \times \left\{ Q_f^2 Q_i^2 -Q_f^2 Q_i T_i - Q_f T_f Q_i^2 + \frac{1}{2} \left(Q_f^2 T_i^2 + T_f^2 Q_i^2 + Q_f T_f Q_i T_i \right) \right.  \\
&& \left. \hspace{0.44cm} +\  \frac{\MZ^2}{\Mll^2}  \times \left[Q_f^2 Q_i T_i + Q_f T_f Q_i^2 + \frac{1}{2} Q_f T_f Q_i T_i  - Q_f^2 Q_i^2 \left( 2 - \frac{\MZ^2 + \Gamma_Z^2}{\Mll^2} \right) \right] \right\} \nonumber \\
&& \hspace{0.46cm} + \  h_2  \times  Q_f T_f Q_i T_i \left( 1+ \frac{\MZ^2}{\Mll^2} \right) \nonumber \\
c_3 &=& h_1 \times  \frac{\MZ^2}{\Mll^2} \times \left[ -Q_f T_f Q_i^2 - Q_f^2 Q_i T_i + 2 Q_f^2 Q_i^2 \left( 1-\frac{\MZ^2 + \Gamma_Z^2}{\Mll^2} \right) \right]  \\
c_4 &=& h_1 \times \frac{\MZ^2}{\Mll^2} \frac{\MZ^2 + \Gamma_Z^2}{\Mll^2} \times Q_f^2 Q_i^2 
\eeq
Explicit mass dependence in each term has been isolated in order to illustrate the cross section's simple UV properties.  In particular we see that only terms $c_{0,1,2}$ contribute when $\Mll \to \infty$, indicating again the fact that the cross section is quadratic in $\hat s_W^2$ at high scales.

\section{Details on Theory Predictions and Uncertainties}
\label{app:thy}

To evaluate the QCD scale and PDF uncertainties, we must evaluate the NNLO cross section.  The runtime of the \texttt{DYNNLO} and \texttt{FEWZ} generators is significant due to the complex fixed order calculation and the fact that we cover a wide swath of phase space for multiple processes and multiple center of mass energies.  Therefore, we employ a numerical technique to more efficiently obtain stable results.  

Consider the PDF uncertainty, given by the standard deviation of the ensemble of PDF set members.  If each PDF variation is evaluated separately (as is the case with \texttt{DYNNLO}), then it is clear that the statistical uncertainties on the results for each variation must be far below the size of the PDF uncertainty to reliably resolve it, which can make runtimes nearly prohibitive given the large number of PDF set members.

\begin{figure}[t]
\begin{center}
\includegraphics[height=7.5cm]{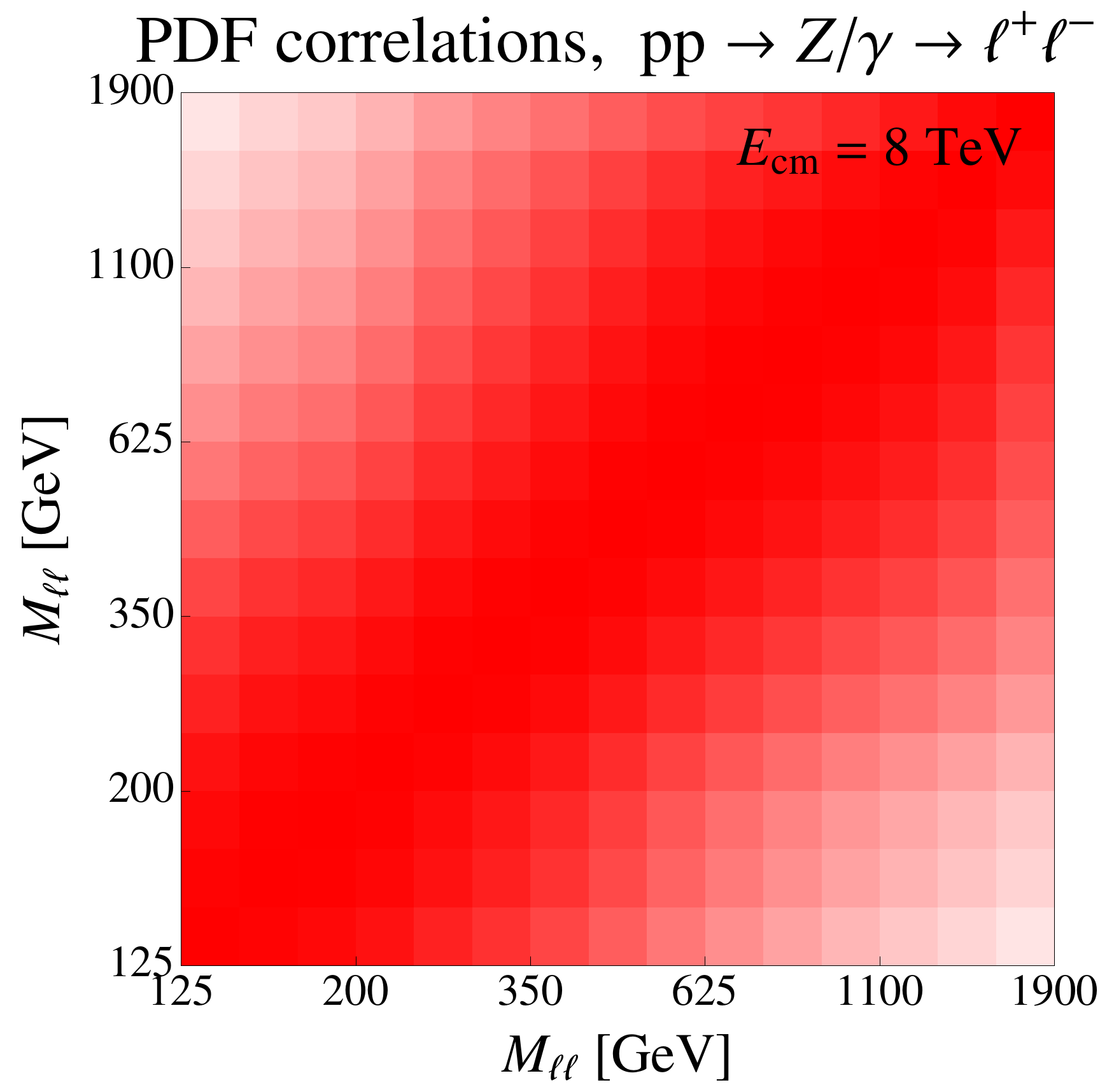}
\includegraphics[height=7.5cm]{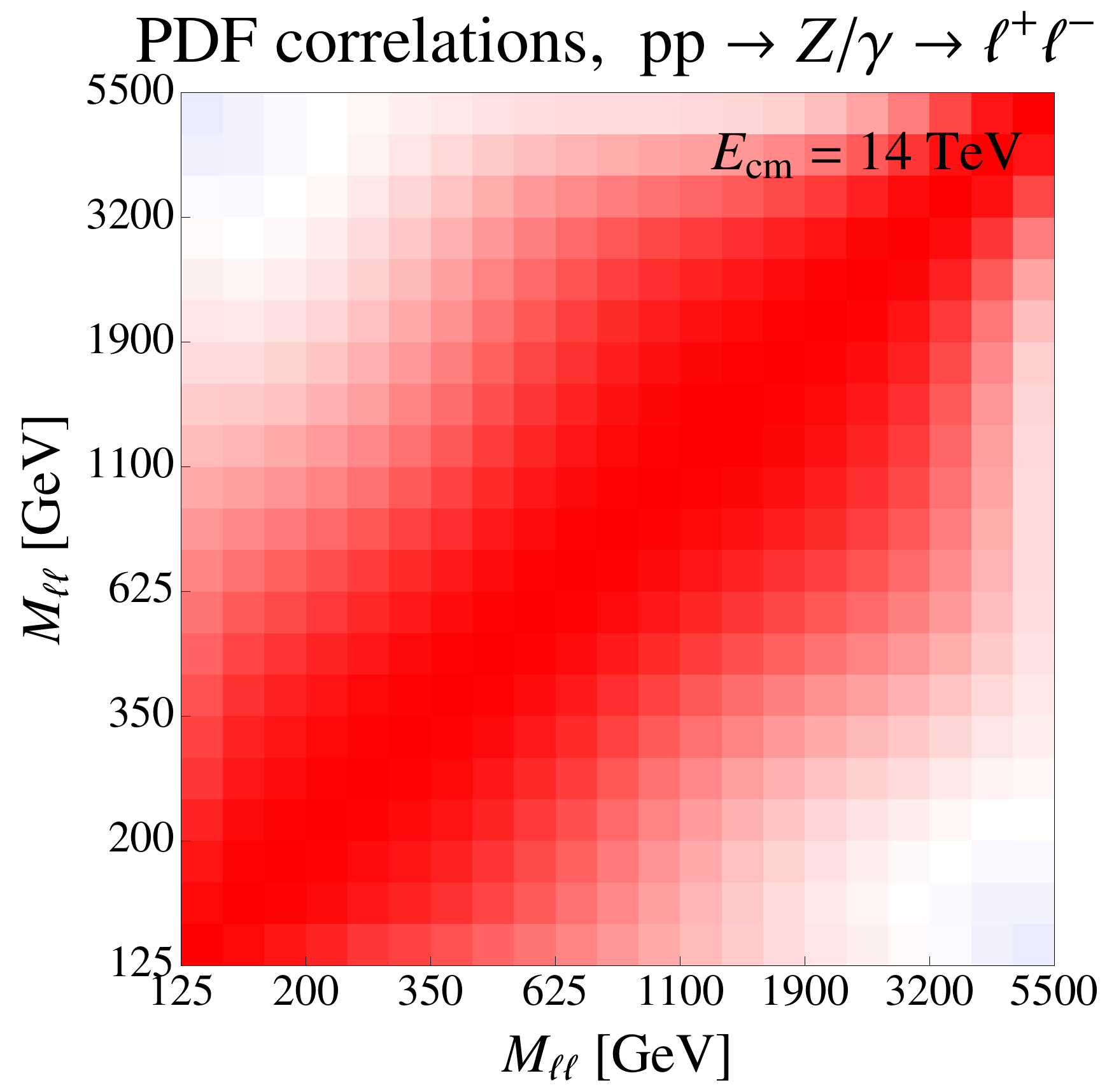}
\includegraphics[height=7.5cm]{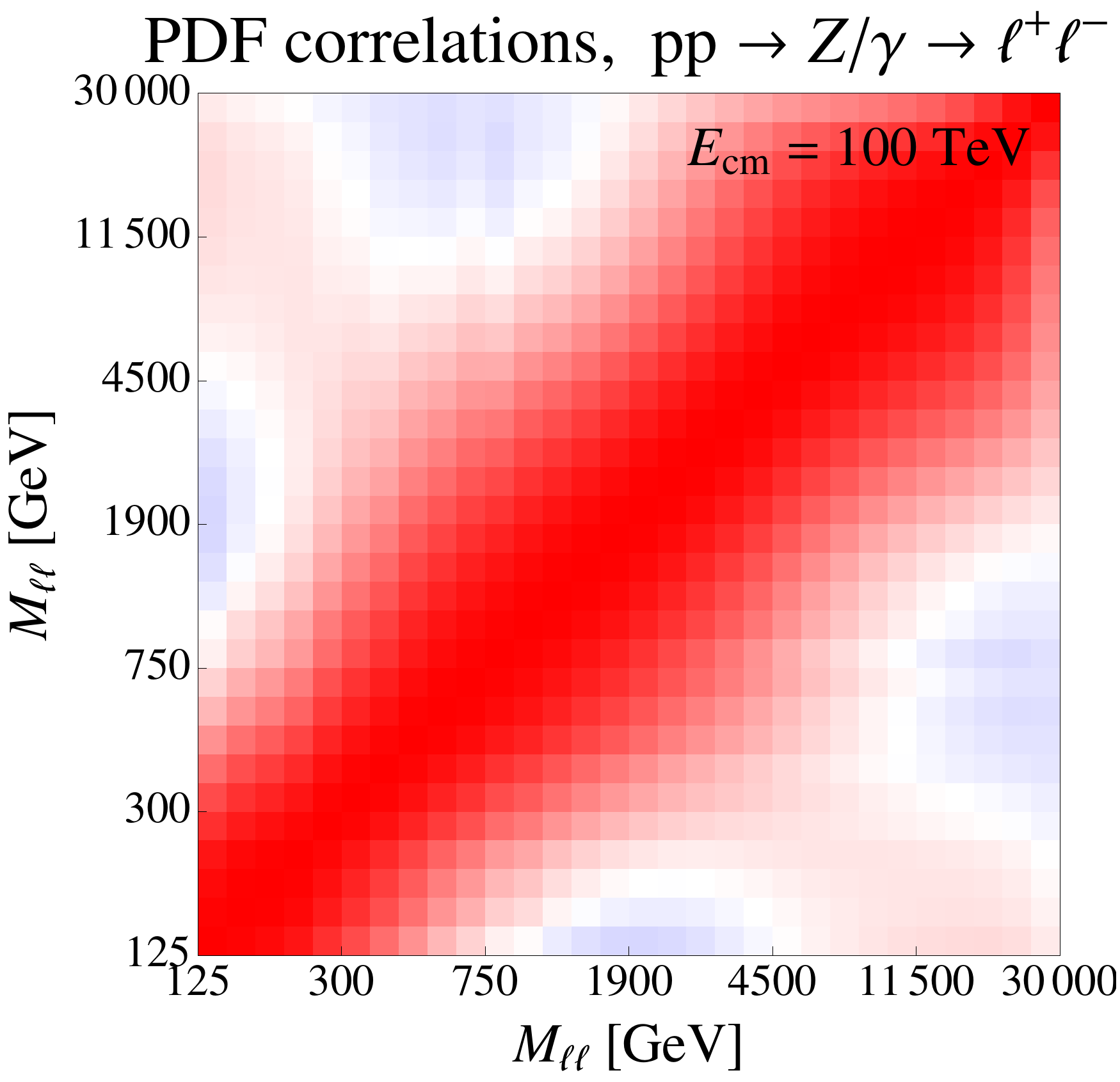}
\raisebox{1.8\height}{\includegraphics[height=1.75cm]{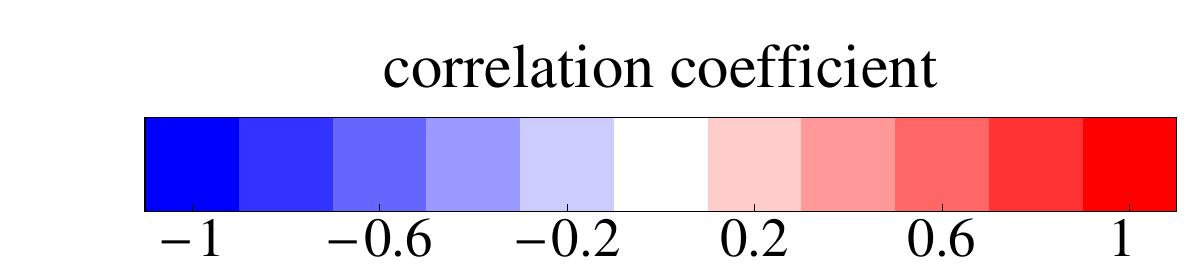}}
\caption{\small{The PDF correlations between bins of $\Mll$ for $\Zproc$ production, plotted for different center of mass energies.}}
\label{fig:DYpdfcorrel}
\end{center}
\end{figure}

Instead of generating fully independent runs, we determine integration grids using the central scale/PDF only and use the grids to obtain results for all scale and PDF variations, correlating the random number seeds among variations.  This ensures that all variations are sampled coherently; for example, a random point in phase space that generates a large weight will be seen by all variations and (largely) factor out of the determination of the uncertainties.  Due to the different matrix elements or PDFs for different variations, the random sample of phase space points still generates statistical fluctuations, but correlating these points between variations removes the statistical fluctuations between different runs of the program evaluating the cross section for different variations.  This largely decouples the statistical uncertainty from the scale and PDF uncertainties, and so we can reliably extract these uncertainties with much less runtime.

A notable feature in the PDF uncertainties is the dependence on the matrix elements.  We find that using the NNLO PDF with either the NNLO matrix elements or the LO matrix elements produces essentially the same \textit{fractional} PDF uncertainties across all three channels ($Z/\gamma$, $W^+$, and $W^-$), collider energies, and $\Mll$ or $\MT$ values.  This is beneficial as the LO matrix elements are tremendously faster to evaluate.  This is likely due to the fact that new parton channels (such as $gg \to \ell^+\ell^-$) do not make a significant contribution to the cross section, and hence the fractional uncertainties are consistent across matrix elements.

In addition to uncertainties in individual bins, it is important to determine the correlations in theory uncertainties between bins.  The QCD scale uncertainties are fully correlated between bins, as a single parameter ($\mu = \mu_R = \mu_F$) determines the uncertainty in all bins, even for different collider energies\footnote{Since all processes are $s$-channel, where the relevant scale for the couplings is $\hat{s}$, we take the scale uncertainties between processes as fully correlated for the purpose of determining a combined limit.}.  This implies that the covariance matrix of QCD scale uncertainties is given by the outer product of the uncertainty amplitudes in each bin with itself (analogously, the correlation matrix has 1 in every entry).  The correlations between bins for the PDF uncertainty may be computed in the usual way,
\begin{equation}
\text{corr} ( \text{bin $i$, bin $j$} ) = \frac{ \sum_k ( \sigma^{(k)}_i - \bar{\sigma}_i )( \sigma^{(k)}_j - \bar{\sigma}_j ) }{ \sum_k ( \sigma^{(k)}_i - \bar{\sigma}_i )^2 \sum_k ( \sigma^{(k)}_j - \bar{\sigma}_j )^2 }
\end{equation}
where $\bar{\sigma}_i$ and $\sigma^{(k)}_i$ are the cross sections in bin $i$ for the central PDF and the $k^{\rm th}$ PDF variation respectively.  In \fig{DYpdfcorrel} we plot these correlations between bins of $\Mll$ for $\Zproc$ production, for various collider energies.  As expected, the correlations are largely near 1, except for bins widely separated in $\Mll$.

\section{Limit Estimation and Statistics}
\label{app:stats}

For the estimation of limits and projected sensitivity on the running of electroweak couplings, we use a profile likelihood test~\cite{Cowan:2011}. Let $\sigma_i^{\text{obs}}$ be the measured Drell-Yan cross section in the {\it i}-th bin, and $\sigma_i^{\text{pred}}(\boldsymbol{\mu,\theta})=\sigma_i^{\text{pred}}(\alpha_1,\alpha_2,\boldsymbol{\theta})$ be the predicted Drell-Yan cross section in the {\it i}-th bin given $\alpha_1(Q)$, $\alpha_2(Q)$ and nuisance parameters $\boldsymbol\theta$. Here, $\boldsymbol{\mu}$ are model parameters that control the running of $\alpha_1(Q)$, $\alpha_2(Q)$. For instance, in a BSM model with an EW-charged state with mass $M$, $\boldsymbol{\mu}$ is given by $\boldsymbol{\mu}=\{M,\Delta b_1,\Delta b_2\}$, where $\Delta b_1$ and $\Delta b_2$ are the contribution to the EW beta function coefficients for $Q>M$. Moreover, let $\Sigma_{ij}$ be the covariance matrix encoding the (theoretical and experimental) uncertainties and their correlations for all bins.

The log-likelihood of a model predicting $\sigma_i^{\text{pred}}(\boldsymbol{\mu,\theta})$ given the observed $\sigma_i^{\text{obs}}$ is:
\begin{equation}
\log \mathcal{L}(\boldsymbol{\mu,\theta}) =\sum_{i,j} -\frac{1}{2} \left(\sigma_i^{\text{pred}}(\boldsymbol{\mu,\theta})-\sigma_i^{\text{obs}}\right).\Sigma_{ij}^{-1}.\left(\sigma_j^{\text{pred}}(\boldsymbol{\mu,\theta})-\sigma_j^{\text{obs}}\right).
\end{equation}
The profiled log-likelihood is given by:
\begin{equation}
\log \lambda(\boldsymbol{\mu}) = \log \mathcal{L}(\boldsymbol{\mu,\hat{\hat{\theta}}}) - \log \mathcal{L}(\boldsymbol{\hat{\mu},\hat{\theta}}),
\end{equation}
where the unconditional maximum-likelihood estimators $\boldsymbol{\hat{\mu},\hat{\theta}}$ maximize $\text{log }\mathcal{L}(\boldsymbol{\mu,\theta})$, while the conditional maximum-likelihood estimator $\boldsymbol{\hat{\hat{\theta}}}$ maximizes $\text{log }\mathcal{L}(\boldsymbol{\mu_0,\theta})$ for a given $\boldsymbol{\mu_0}$. If the model described by $\boldsymbol{\mu}$ has $n$ independent degrees of freedom, then the $p$-value for this model given the observations is determined by 
\begin{equation}
p=1-\text{CDF}(\chi^2_n,-2~\log \lambda(\boldsymbol{\mu})),
\end{equation}
where $\text{CDF}(\chi^2_n,-2~\text{log }\lambda(\boldsymbol{\mu}))$ is the cumulative distribution function for a $\chi^2$-distribution with $n$ degrees of freedom evaluated at $-2~\log \lambda(\boldsymbol{\mu})$.  From the $p$-value we assess  the various statistical significances quoted in \sec{limits}.

For the 7~TeV measurements performed by ATLAS and CMS, we build the covariance matrix $\Sigma_{ij}$ as a sum of experimental uncertainties (provided in~\cite{ATLAShepdata,CMShepdata}) and theoretical (PDF and scale) uncertainties:
\begin{equation}
\Sigma_{ij}^{\text{ATLAS}}=\Sigma_{ij}^{\text{exp, ATLAS}}+\Sigma_{ij}^{\text{PDF}}+\Sigma_{ij}^{\text{scale}},
\end{equation}
and likewise for CMS\@.

For projected sensitivities at 8,~14,~and 100~TeV, we assume $\sigma_i^{\text{obs}}$ is consistent with the NNLO QCD prediction in the Standard Model, and marginalize over the normalization of the $M_{\ell\ell}$ ($M_T$) distribution to factor out unknown correlated systematics ({\it i.e.}, we treat the Drell-Yan normalization as a nuisance parameter $\boldsymbol\theta$). When combining different $\sqrt{s}$, such as 8,~14~and~100~TeV, we take separate normalizations for each $\sqrt{s}$. We build the covariance matrix as: 
\begin{equation}
\Sigma_{ij}=\Sigma_{ij}^{\text{DY stat}}+\Sigma_{ij}^{\text{uncorr sys}}+\Sigma_{ij}^{\text{PDF}}+\Sigma_{ij}^{\text{scale}},
\end{equation}
where the diagonal matrices $\Sigma_{ij}^{\text{DY stat}}$, $\Sigma_{ij}^{\text{uncorr sys}}$ encode the uncorrelated statistical and systematic experimental uncertainties in the Drell-Yan cross section measurements. Unless otherwise specified, $\Sigma_{ij}^{\text{uncorr sys}}$ is built assuming a flat 1\% uncertainty across all invariant (transverse) mass bins.

\section{Effects on Limits from the Scale Choice for Electroweak Couplings}
\label{app:EWscale}

Limits on new physics from measurements of running EW couplings depend on the scale dependence of the couplings in the Drell-Yan cross sections.  For example, we have made the central scale choices of $\Mll$ and $\Mlnu$ for the neutral and charged current cross sections, respectively (see \eqs{sigmaZgamma}{sigmaW}).  Since the processes are $s$-channel, and this cross section contains logarithms of the form $\ln M^2 / M_{\rm EW}^2$, where $M = \Mll, \Mlnu$ and $M_{\rm EW} = M_Z, M_W$, this choice of the central scale is robust, and resummation of the EW Sudakov logarithms should produce rapid convergence of the cross section in resummed perturbation theory, with small uncertainties.  A standard variation of the scale between $\Mll/2$, $\Mlnu/2$ and $2\Mll$, $2\Mlnu$ provides an uncertainty on the prediction.  This affects the limit produced, and this effect is part of the EW uncertainty, which is part of the global uncertainty budget.  However, a precise calculation of the EW effects, one that includes the NLO terms and may require resummation of virtual and real corrections, will contribute a small uncertainty relative to other uncertainties, and we have therefore not included the EW scale uncertainty in setting the limit.  As our limit procedure only uses the LO EW dependence to compute limits, it is particularly sensitive to the scale choice.

\begin{figure}[t]
\begin{center}
\includegraphics[width=\textwidth]{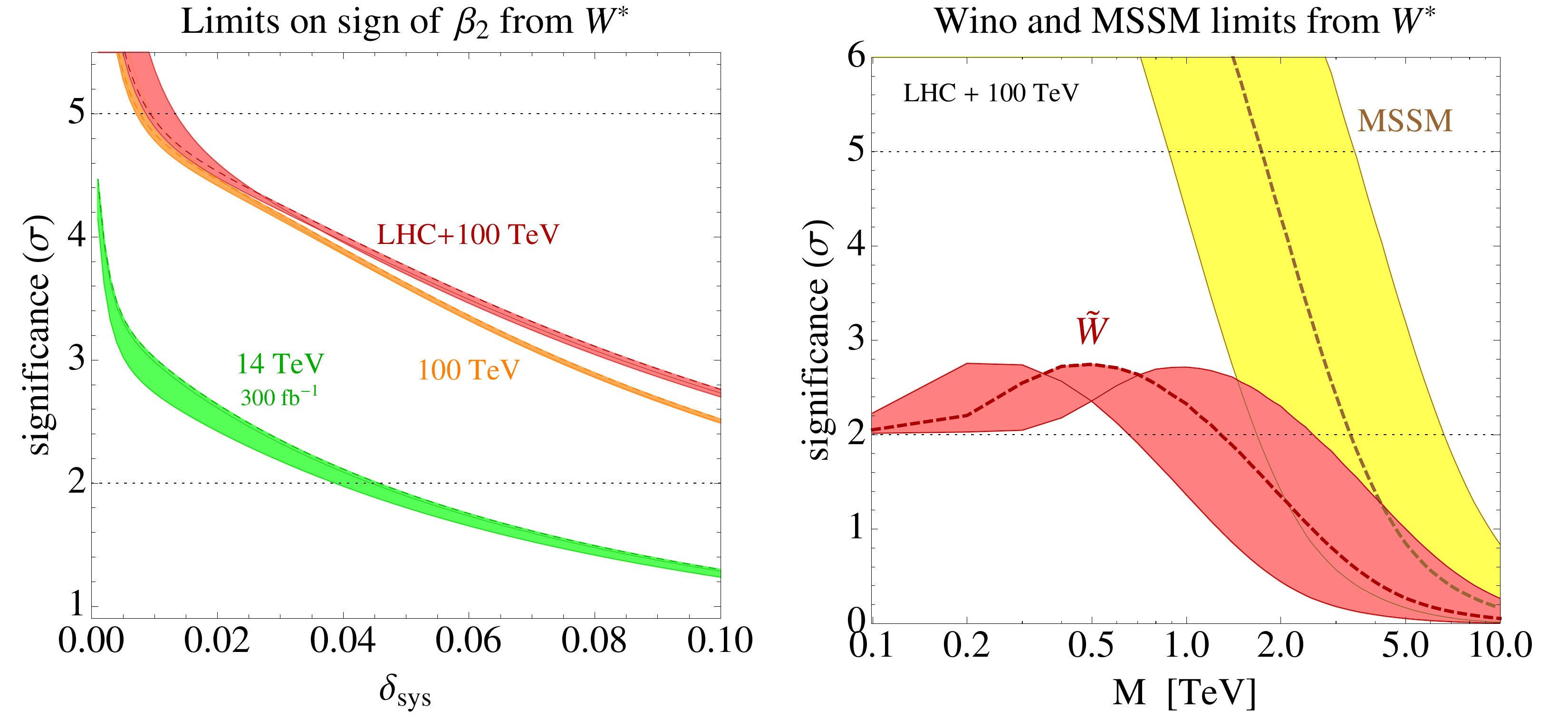}
\caption{\small{The reach for the determination of the sign of $\beta_2$ as a function of the size of the uncorrelated systematic uncertainties (left plot) and the reach for $\tilde W$ and the MSSM as a function of mass $M$ (right plot) from $W^*$.  The reach is shown for three different EW coupling scale choices: $\Mlnu$, $\Mlnu/2$, and $2\Mlnu$.  The EW scale dependence is only used at LO, which leads to large uncertainties in the reach; a precision EW calculation will tighten the band.}}
\label{fig:reachBands}
\end{center}
\end{figure}

We find it useful, therefore, to show the variation in the reach as the scale choice for the EW coupling is varied.   This variation is conservative, and will be reduced by making robust predictions for the Drell-Yan cross sections that include higher order EW effects from SM and BSM matter.  In \fig{reachBands} we show the reach for a wino and the MSSM as a function of the mass scale $M$, and the determination of the sign of $\beta_2$ as a function of the size of the uncorrelated systematic uncertainty, using the $W^*$ channel.  The reach is shown for different scale choices ($\Mlnu $, $\Mlnu/2$, and $2\Mlnu$) of the LO EW coupling.  The determination of the sign of $\beta_2$ is largely unaffected by the scale choice, even using only the LO EW dependence.  For a given significance, the width of the reach band for a wino and the MSSM is approximately a factor of 4, which corresponds to the overall factor of 4 in the scale variation.  The size of the band will significantly decrease when the EW corrections are included to higher orders.  This motivates the inclusion of BSM matter content into the calculation of higher order EW corrections, particularly in public codes that currently compute the analogous corrections in the SM.


\end{document}